\documentclass[12pt]{article}
\pdfoutput=1 

\usepackage{cancel,slashed}
\usepackage{bbm}
\usepackage{amssymb}
\usepackage{amsfonts}
\usepackage{amsmath}
\usepackage{graphicx}
\usepackage{cite}

\usepackage{latexsym}
\usepackage{color}
\usepackage{xypic}
\usepackage{cancel,slashed}
\usepackage[linktocpage]{hyperref}
\usepackage{color}
\usepackage{transparent}
\usepackage{footmisc}

\makeatletter
\def\@xfootnote[#1]{%
  \protected@xdef\@thefnmark{#1}%
  \@footnotemark\@footnotetext}
\makeatother

\newcommand{\bmat}{\left(\begin{array}}
\newcommand{\emat}{\end{array}\right)}

\def\p{\partial}
\def\a{\alpha}

\def\b{\beta}

\def\th{\theta}

\def\-{\hphantom{-}}

\def\s2{\frac{1}{\sqrt2}}

\def\oh{\frac{1}{2}}
\def\beq{\begin{equation}}
\def\eeq{\end{equation}}
\def\beqa{\begin{eqnarray}}
\def\eeqa{\end{eqnarray}}

\def\tr{{\rm tr \,}}

\def\Dsl{\,\raise.15ex\hbox{/}\mkern-13.5mu D} 

\def\CM {{\cal M}}

\def\CN {{\cal N}}

\def\CL {{\cal L}}

\def\CQ {{\cal Q}}

\def\tr{\mbox{Tr}}

\def\be{\begin{equation}}
\def\ee{\end{equation}}
\def\bea{\begin{eqnarray}}
\def\eea{\end{eqnarray}}
\def\raw{\rightarrow}

\def\IR{\mathbb{R}}

\def\Id{{\mathbb{I}}}

\def\oh{\frac{1}{2}}
\def\a{{\alpha}}
\def\b{{\beta}}
\def\de{{\delta}}
\def\eps{{\epsilon}}
\def\th{{\theta}}
\def\Lam{{\Lambda}}
\def\lam{{\lambda}}

\def\sig{{\sigma}}
\def\Sig{{\Sigma}}

\def\G{{\Gamma}}

\def\p{{\partial}}

\makeatletter
\newsavebox{\@brx}
\newcommand{\llangle}[1][]{\savebox{\@brx}{\(\m@th{#1\langle}\)}%
  \mathopen{\copy\@brx\kern-0.5\wd\@brx\usebox{\@brx}}}
\newcommand{\rrangle}[1][]{\savebox{\@brx}{\(\m@th{#1\rangle}\)}%
  \mathclose{\copy\@brx\kern-0.5\wd\@brx\usebox{\@brx}}}
\makeatother

\def\w{{\wedge}}

\newcommand{\conj}{\overline}
\newcommand{\mr}{\mathrm}
\newcommand{\mc}{\mathcal}
\newcommand{\mf}{\mathbf}
\newcommand{\mb}{\mathbb}
\newcommand{\mg}{\mathfrak}
\newcommand{\pb}{\conj{\p}}

\def\sm2{{\mbox{\small 2}}}

\newcommand{\bp}{\begin{pmatrix*}[r]}  
\newcommand{\ep}{\end{pmatrix*}}  
\newcommand{\bpp}{\begin{pmatrix}}  
\newcommand{\epp}{\end{pmatrix}}  
\newcommand{\bcd}{\begin{center}
\begin{tikzcd}}
\newcommand{\ecd}{\end{tikzcd} \end{center}}

\def\P{\mathbb{P}}
\def\cL{\mathcal{L}}

\def\1{\mathbb{1}}

\def\h{{\rm h}}

\def\del{\partial}
\def\delbar{\bar{\partial}}

\def\L{\mathcal{L}}

\def\M{\mathcal{M}}
\def\P{\mathcal{P}}

\begin{document}
\pagestyle{plain}

\makeatletter
\@addtoreset{equation}{section}
\makeatother
\renewcommand{\theequation}{\thesection.\arabic{equation}}
\pagestyle{empty}
\rightline{ IFT-UAM/CSIC-19-015} \rightline{ROM2F/2019/01}
\vspace{0.5cm}
\begin{center}
\Huge{{T-branes and defects}
\\[15mm]}
\normalsize{Fernando Marchesano,$^1$ Raffaele Savelli,$^{2}$ and Sebastian Schwieger$^1$ \\[10mm]}
\small{
${}^1$ Instituto de F\'{\i}sica Te\'orica UAM-CSIC, Cantoblanco, 28049 Madrid, Spain \\[2mm] 
${}^2$ Dipartimento di Fisica, Universit\`a di Roma ``Tor Vergata" \& INFN - Sezione di Roma2 \\
Via della Ricerca Scientifica 1, 00133 Roma, ITALY
\\[8mm]} 
\small{\bf Abstract} \\[5mm]
\end{center}
\begin{center}
\begin{minipage}[h]{15.0cm}

We study T-branes on compact K\"ahler surfaces, in the presence of fields localised at curves. If such fields are treated as defects, their vevs induce delta-function sources for the 7-brane background, possibly leading to profiles with poles. We find that the presence of defect sources relaxes the constraints on globally well-defined T-brane configurations, avoiding the obstruction to building them on surfaces of positive curvature. Profiles with poles can be understood, from a 4d viewpoint, as non-trivial vevs for massive modes induced by the defects, and come with their own set of constraints. In the special case of fields localised on a self-intersection curve, we show how to retrieve the Hitchin system with defects from an ordinary global one with enhanced symmetry.

\end{minipage}
\end{center}
\newpage
\setcounter{page}{1}
\pagestyle{plain}
\renewcommand{\thefootnote}{\arabic{footnote}}
\setcounter{footnote}{0}


\tableofcontents


\section{Introduction}
\label{s:intro}

One of the open problems in string theory is to characterise the spectrum of extended BPS objects of a given string compactification. This question remains unanswered even for particularly well-studied cases of vacua, like BPS D-branes in type IIB Calabi-Yau orientifolds with O3/O7-planes, or their F-theory generalisation \cite{thebook,Denef:2008wq,Heckman:2010bq,Weigand:2010wm,Wijnholt:2012fx,Maharana:2012tu,Weigand:2018rez}. In this case one can construct an abundant set of space-time-filling BPS D-branes at large volume, by  considering D3-branes located at points in the internal six-manifold $B$, or 7-branes wrapping holomorphic four-cycles $S \subset B$. One may even enlarge this set by endowing the 7-branes with anti-self-dual gauge bundles threading $S$. In some instances, these can be understood as bound states of 7-branes and D3-branes, as they carry the same charges. 

In general, classifying the possible BPS bound states of D-branes at a given point in closed string moduli space is where a large fraction of the problem resides. One particular set of such bound states, that have recently been the subject of attention, are stacks of 7-branes whose complex worldvolume scalar or Higgs field has an intrinsic non-Abelian profile. These objects, typically dubbed Higgs bundles or T-branes in the string theory literature \cite{Donagi:2003hh,Hayashi:2009bt,Cecotti:2010bp,Donagi:2011jy}, can be seen as generalisations of the original construction of Hitchin \cite{Hitchin:1986vp}. As pointed out in \cite{Hayashi:2009bt,Cecotti:2010bp}, such T-brane backgrounds are crucial to engineer and compute realistic Yukawas in F-theory GUTs \cite{Donagi:2008ca,Beasley:2008dc,Beasley:2008kw,Donagi:2008kj}, and have also been proposed as a key ingredient to engineer de Sitter vacua \cite{Cicoli:2015ylx}. Since then, there has been a lot of effort to understand these BPS objects and their r\^ ole in string compactifications \cite{Chiou:2011js,Donagi:2011dv,Font:2013ida,Anderson:2013rka,DelZotto:2014hpa,Collinucci:2014qfa,Collinucci:2014taa,Marchesano:2015dfa,Carta:2015eoh,Heckman:2016ssk,Collinucci:2016hpz,Bena:2016oqr,Marchesano:2016cqg,Mekareeya:2016yal,Ashfaque:2017iog,Anderson:2017rpr,Bena:2017jhm,Collinucci:2017bwv,Cicoli:2017shd,Marchesano:2017kke,Anderson:2017zfm,Apruzzi:2018oge,Cvetic:2018xaq,Heckman:2018pqx,Apruzzi:2018xkw,Carta:2018qke}. 

In \cite{Marchesano:2017kke} we initiated the study of T-branes from a global perspective, determining the conditions for BPS T-branes to exist on a compact K\"ahler surface $S$. Remarkably, we found stringent constraints on the gauge bundles defining the T-brane background, as well as on the canonical bundle of $S$. In particular, under the assumption of Abelian gauge bundles and Higgs fields without poles, we found an obstruction to building globally well-defined T-brane configurations in surfaces of positive or vanishing Ricci curvature. 

The analysis in \cite{Marchesano:2017kke} also assumed that the 7-brane theory wrapping $S$ can be isolated from the other sectors of the compactification. More precisely, it assumed that the 7-brane BPS equations do not depend on the presence of other objects, like further 7-branes wrapping other complex surfaces. Fully-fledge compactifications will contain several stacks of 7-branes wrapping different holomorphic four-cycles, and generically some of them will  have non-trivial intersections with the surface $S$ hosting the T-brane background. However, due to the no-force condition between BPS objects, one may safely ignore their presence when analysing the BPS equations for the 7-brane background on $S$. 

This statement no longer holds if one turns on the degrees of freedom charged under different 7-brane sectors. In particular, one may consider those localised at the intersection of two 7-brane stacks. If non-trivial vevs are given to the fields at $\Sig = S \cap S'$, then the previous BPS conditions for the 8d theories on $S$ and $S'$ are modified, and the two sectors of 7-branes are bound to each other. Particularly interesting are those bound states for which the locus $S \cup S'$ remains geometrically unchanged, which can also be thought of as a kind of T-brane \cite{Donagi:2011jy,Collinucci:2014qfa}. With this procedure one may thus construct even more involved examples of bounded BPS objects, enlarging the previous set of 7-brane configurations. 

In this paper we implement this strategy to extend the analysis of \cite{Marchesano:2017kke}, and  consider T-brane configurations on $S$ bound to a second stack of 7-branes on $S'$. Following \cite{Beasley:2008dc}, one may use the localisation properties of 7-brane intersections to treat the degrees of freedom on $\Sig = S \cap S'$ as a defect 6d theory on $\IR^{1,3} \times \Sig$ coupled to the 8d theory on $\IR^{1,3} \times S$. As such, one can describe the effect of forming the bound state as a deformation of the BPS 7-brane equations on $S$ which, as we turn on vevs for the defect fields, acquire source terms with $\delta$-function support at $\Sig$. In practice, this means that T-brane profiles can now have poles along $\Sig$, or in other words that the T-brane spectrum is enlarged to comprise configurations built out of both holomorphic and meromorphic sections on $S$.\footnote{Hitchin-like systems coupled to defects have already been considered within the T-brane literature \cite{Anderson:2013rka,Anderson:2017zfm} in the context of compactifications to six dimensions. Though with different uses and motivations, they also play a key r\^ole in the study of the so-called field theories of class $\mathcal{S}$ \cite{Gaiotto:2009hg}.}

One particular motivation for this analysis is to see if the previous no-go result to building T-branes on surfaces of positive and vanishing Ricci curvature holds for this more general class of configurations. We find that, in the presence of defect sources, the global constraints found previously are substantially relaxed, and that one can indeed consider $S$ to be, e.g., a del Pezzo surface. This can be achieved with both regular and singular Higgs-field profiles along $\Sig$. The key feature of these constructions is that they require the existence of defect zero modes instead of bulk zero modes, and this is a much less stringent condition from the viewpoint of the topology of $S$ and the gauge bundle. We will in fact see that, in our setup, meromorphic Higgs profiles do not correspond to a new degree of freedom to build T-brane configurations, but that they are instead determined by the defect-field vevs. More precisely, by dimensional reduction to 4d, we find that a pair of defect modes can have Yukawa couplings with a whole tower of bulk KK modes. Giving this pair a vev in turn induces a vev for the massive modes of the Higgs field, such that the collective effect is the appearance of a pole in its 8d profile. Besides a shift on the effective Fayet-Iliopoulos term, there is no other effect of the meromorphic Higgs profile at low energies. What remains at this level are the Yukawa couplings of this pair of defect fields with the massless modes of the tower. This may obstruct switching on such defect fields and creating the corresponding 7-brane bound state. 

One particularly interesting setup in this framework is when the intersecting surface $S'$ is a holomorphic deformation of $S$, and the defect locus $\Sig$ is the self-intersecting curve of the surface $S$ hosting the T-brane. In this case, if the deformation is small in string units, one should be able to describe the whole 7-brane configuration as a regular Hitchin system with enhanced symmetry group $G_\Sig \supset G_S \times G_{S'}$, as opposed to two Hitchin systems with symmetry groups $G_S$ and $G_{S'}$, respectively, coupled to the defect sector. We will consider such a regular Hitchin system and work out the dictionary to the one with defects. In particular, we will see how the BPS equations of the former reproduce those of the latter upon taking appropriate limits, recovering the $\de$-function sources characteristic of defects.

The paper is organised as follows. In section \ref{s:defects} we briefly review the results of \cite{Marchesano:2017kke} and extend them by adding defect sources. We devise two different schemes in which the bulk T-brane background can be hosted by a surface of positive curvature, and dub them holomorphic and meromorphic schemes. In section \ref{s:4d} we discuss the defect schemes from another perspective, that of the 4d effective theory.  In particular, we show that the pole that appears in the meromorphic scheme can be understood as a series of massive KK modes acquiring a vev due to the Yukawa couplings developed between bulk and defect modes. We also discuss how similar Yukawas with the massless bulk sector may provide an obstruction to realising such a configuration. In section \ref{s:Hitchin} we analyse both of these schemes in the particular case of two intersecting homotopic surfaces, which allows to describe the degrees of freedom of the whole configuration in terms of a Hitchin system without defects. We show the two different limits in which one can recover the BPS equations of the corresponding defect schemes, and discuss the range of validity for each of these two descriptions.  We draw our conclusions in section \ref{s:conclu}.

Some technical details have been relegated to the appendices. In appendix \ref{ap:su3} we specify our Lie algebra conventions used throughout the main text. Appendix \ref{ap:defects} presents the general BPS equations in the presence of defects.  Finally, Appendix \ref{ap:4d} contains all the details of the dimensional reduction used in section \ref{s:4d}.


\section{T-branes with defects}
\label{s:defects}

In this section we discuss the existence of T-brane backgrounds which are globally well-defined on a compact complex surface $S$, reviewing and extending the analysis of \cite{Marchesano:2017kke}. The class of T-brane backgrounds analysed in \cite{Marchesano:2017kke} are those in which the 7-brane theory on $S$ is isolated from the other 7-branes of the compactification, in the sense that the presence of the latter does not affect the BPS conditions of the former. In the following we will generalise this setup, by allowing for vevs of fields localised at the intersection of $S$ and a four-cycle $S'$ wrapped by a second 7-brane stack. Such vevs imply that both 7-brane stacks form a bound state, and in particular that their BPS conditions are coupled. This generalisation can be understood as a modification of the previous BPS conditions, if the fields at the intersection $\Sig = S \cap S'$ are treated as a defect theory coupled to the 7-brane theory on $S$ \cite{Beasley:2008dc}. We will take such approach to classify this more general class of T-brane backgrounds, and we will see that the obstruction found in \cite{Marchesano:2017kke} to building T-branes on compact surfaces is relaxed in the presence of defects. In particular, it is possible to construct such backgrounds when $S$ is a surface of positive curvature, unlike it was in the absence of defects.

\subsection{Compact T-brane systems}
\label{ss:compact}

Let us consider F-theory on $\IR^{1,3} \times \CM$, with $\CM$ a Calabi-Yau four-fold, and a stack of 7-branes wrapping a compact K\"ahler surface $S$ of the three-fold base of $\CM$. In general, the precise 7-brane configuration and its lightest degrees of freedom can be specified by an eight-dimensional super-Yang-Mills theory on $\IR^{1,3} \times S$ with symmetry group $G_S$ \cite{Donagi:2008ca,Beasley:2008dc,Beasley:2008kw,Donagi:2008kj}. The two objects defining such an action are the field strength $\mathbb{F} = d\mathbb{A} - i\mathbb{A} \wedge \mathbb{A}$ of the 7-branes gauge boson $\mathbb{A}$, and the (2,0)-form Higgs field $\Phi$, whose eigenvalues describe the 7-brane transverse geometrical deformations. Both $\mathbb{A}$ and $\Phi$ transform in the adjoint of the symmetry group $G_S$ and, whenever they have a non-trivial profile, they break the gauge group to a subgroup of $G_S$. The 7-brane configurations that preserve supersymmetry correspond to those solving the following set of conditions
\begin{subequations}
\label{susy7}
\begin{align}
\label{Fterm1}
\bar \partial_{\mathbb{A}} \Phi = &\, 0\, ,\\
\label{Fterm2}
\mathbb{F}^{(0,2)} =&\, 0\, ,\\
\label{Dterm}
J \wedge \mathbb{F} +\frac{1}{2} [\Phi, \Phi^\dagger] =&\,0\,,
\end{align}
\end{subequations}
with $J$ the K\"ahler two-form of $S$. From the four-dimensional viewpoint, the first two equations imply vanishing F-terms, while the third one ensures vanishing D-terms.

In \cite{Marchesano:2017kke} we analysed supersymmetric 7-brane backgrounds with non-commuting expectation values for the Higgs field $\Phi$. In other words we considered configurations solving \eqref{susy7} and such that $[\Phi, \Phi^\dagger] \neq 0$, also known as T-branes in the string theory literature \cite{Donagi:2003hh,Hayashi:2010zp,Cecotti:2010bp,Donagi:2011jy}. We imposed that such T-brane configurations are globally well-defined over $S$ and that the Higgs-field profile is absent of poles. In the remainder of this subsection we will review some of the main results obtained in \cite{Marchesano:2017kke}, and in the next one we will see how these results are modified when we allow for the presence of defects, which may induce poles. 

The simplest T-brane configuration that one may construct is based on the symmetry group $G_S = SU(2)$, which in applications to F-theory GUTs one may identify with a subgroup transverse to $G_{\rm GUT}$. In this case, the simplest non-trivial gauge bundle that one may consider on  $S$ is of rank two and split type,\footnote{As in \cite{Marchesano:2017kke}, we will always assume that $S$ is simply-connected, i.e. $\pi_1(S)=0$. This implies that holomorphic line bundles on $S$ have their topology completely specified by their first Chern class.} namely $\mathbb{V}=\mathcal{L}\oplus\mathcal{L}^{-1}$. Due to the BPS equation \eqref{Fterm2}, $\mathcal{L}$ is endowed with a holomorphic structure. Then, if  $\{T_+, T_-, T_3\}$ with $[T_+, T_-] = T_3$, are the generators of $\mathfrak{sl}(2)$ this translates into a flux background of the form $\mathbb{F} = F \, T_3$, which in the fundamental representation reads
\be
\mathbb{F}=\left(\begin{array}{cc}
F&0\\ 0 &-F
\end{array}\right)\, , \qquad \qquad F=F^{\h}-i\del\delbar g\, ,
\label{decompsu2}
\ee
with $F^{\h}$ a harmonic (1,1)-form and $g$ a real function, both globally well-defined on $S$. Fixing the holomorphic structure of $\mathcal{L}$ such that $A^{(0,1)} = 0$, a choice usually dubbed holomorphic gauge \cite{Font:2009gq,Cecotti:2009zf}, allows to rewrite everything in terms of the hermitian metric $h$ on $\L$. In particular locally we can write $F=-i\del\delbar \log h$, with $h=h_0\, e^g$ and $h_0$ the metric that corresponds to $F^{\h}$.

One may pair up this flux background with a nilpotent Higgs background of the form $\Phi = m T_+$, or
\be\label{NilpPhi}
\Phi=\left(\begin{array}{cc}
0& \mf{m} \\ 0 &0
\end{array}\right)\, .
\ee
where, in the holomorphic gauge, $\mf{m} \in H^{2,0}(S,\CL^{2})$. Equivalently, we can also see $\mf{m}$ as a scalar holomorphic section $m$ of the line bundle $\CM \equiv \mathcal{L}^2\otimes K_S$, with $K_S$ the canonical bundle of $S$.\footnote{Throughout the text, boldface quantities like $\mf{m}$ will denote holomorphic (2,0)-forms, while the same letter in italic will stand for their scalar counterpart through the canonical isomorphism $H^{2,0}(\mc{V}) \simeq H^0(K_S \otimes \mc{V})$ for an arbitrary bundle $\mc{V}$.} The existence of $m$ implies that $\CM$ is effective in $S$, and therefore
\be\label{ExistenceOfM}
\int_SJ\wedge c_1(\M)=\int_SJ\wedge(2c_1(\L)+c_1(K_S))\ge0
\ee
with the equality holding if and only if $\M$ is trivial and $m$ constant. On the other hand, the D-term condition \eqref{Dterm} for this background reads
\be
 \int_SJ\wedge c_1(\L) = - \frac{1}{8\pi} \tr \int_S [\Phi,\Phi^\dagger] T_3 < 0 \, .
\ee
Taking both equations into account one obtains the following set of inequalities 
\be
\int_SJ\wedge c_1(K_S)>  \int_SJ\wedge c_1(\M)\geq 0\,,
\label{ineqnil}
\ee
which constrains the viable choices for the bundle $\CL$ and forbids $S$ to be K3 or a manifold with positive-definite Ricci curvature. One may complicate the Higgs-field background and replace \eqref{NilpPhi} by
\be\label{NonNilpotentPhiStack}
\Phi=\left(\begin{array}{cc}
0&\mf{m}\\ \mf{p} &0
\end{array}\right)\,,
\ee
where $\mf{p}\in  H^{2,0}(S,\CL^{-2})$ defines an element of $H^0(S,\P)$, with $\P \equiv \mathcal{L}^{-2}\otimes K_S$. Now $\P$ also needs to be effective and, as discussed in \cite{Marchesano:2017kke}, from the D-term equation one recovers a hierarchy of curve areas that imposes $\int_SJ\wedge c_1(K_S)>0$ and restricts the possible choices for $\CL$. This is consistent with the fact that $mp$ transforms as a section of $H^0(S,K_S^2)$, and as such implies a holomorphic deformation for the surface $S$ that is forbidden in the case of positive curvature manifolds like, e.g., del Pezzo surfaces \cite{Hayashi:2010zp}.

\subsubsection*{The no-go for $SU(3)$}

This no-go result generalises to arbitrary T-brane backgrounds with higher rank gauge group $G$, as long as the worldvolume fluxes lie along its Cartan subalgebra. For concreteness here we will only review the case of $G_S = SU(3)$, 
which will become useful at the end of this section for understanding how the no-go result can fail in the presence of defects. We refer the reader to \cite{Marchesano:2017kke} for the general proof of the no-go theorem. 

Let us consider $\mathbb{F}$ and $\Phi$ taking values in the complexification of the $SU(3)$ algebra, with their profiles expressed in terms of the Chevalley basis $\{\eta_1,\eta_2, \eps_1, \eps_2, \eps_{12}, \th_1, \th_2, \th_{12}\}$ of $\mg{sl}(3)$ (see Appendix \ref{ap:su3} for explicit expressions). By assumption we have a worldvolume flux valued along the Cartan subalgebra. That is 
\be
\mathbb{F} = F_1\, \eta_1 + F_2\, \eta_2\, .
\label{Fsu3}
\ee
In addition, we have a Higgs field profile valued outside of the Cartan subalgebra,\footnote{Notice that deformations along the Cartan subalgebra are forbidden in positive curvature manifolds.} but such that the commutator $[\Phi, \Phi^\dagger]$ lies within it in order to satisfy the D-term equations. This condition restricts the possible profiles for $\Phi$, allowing it to have non-vanishing components only up to three independent roots. For instance, one may consider the profile $\Phi = m_1 \eta_1 + m_2 \eta_2 + p_{12} f_{12}$. That is, in the fundamental representation of $\mg{sl}(3)$ we have the profiles
\be
\mathbb{F} = 
\left(\begin{array}{ccc}
F_1&0 &0 \\ 0 & F_2 - F_1& 0 \\ 0 & 0 & -F_2
\end{array}\right)\, , \quad \quad 
\Phi = \left(\begin{array}{ccc}
0 & \mf{m}_1&0 \\ 0 & 0 & \mf{m}_2 \\ \mf{p}_{12} & 0 & 0
\end{array}\right)\, ,
\label{FPhisu3}
\ee
where $F_{1,2}$ are closed (1,1)-forms such that $[F_i] = 2\pi c_1(\CL_i)$ and, in the holomorphic gauge, $\mf{m}_1 \in H^{2,0}(\CL_1^2 \otimes \CL_2^{-1})$,  $\mf{m}_2 \in H^{2,0}(\CL_1^{-1} \otimes \CL_2^{2})$, $\mf{p}_{12} \in H^{2,0}(\CL_1^{-1} \otimes \CL_2^{-1})$, with some of these sections possibly vanishing. The D-term equation implies that\footnote{Given a split bundle metric $H_{\mg{su}(3)}  = {\rm diag} (h_1, h_1^{-1} h_2, h_2^{-1})$, we define $\| m_1 \|^2 = \int_S h_1^2h_2^{-1} \mf{m}_1 \wedge \bar{\mf{m}}_1$, $\| m_2 \|^2 = \int_S h_1^{-1} h_2^{2} \mf{m}_2 \wedge \bar{\mf{m}}_2$ and $\| p_{12} \|^2 = \int_S h_1^{-1} h_2^{-1} \mf{p}_{12} \wedge \bar{\mf{p}}_{12}$.}
\be
4\pi \int_S c_1(\CL_1) \wedge J = - \| m_1 \|^2 + \| p_{12} \|^2\, , \quad \quad 4\pi \int_S c_1(\CL_2) \wedge J = - \| m_2 \|^2 + \| p_{12} \|^2\, .
\label{dtermsu3}
\ee
In addition, each non-vanishing holomorphic section implies an effectiveness constraint which, in turn, implies a positivity condition at any point in the K\"ahler cone:
\begin{subequations}
\label{effsu3}
\begin{align}
\label{m1}
m_1 \, \raw \ & A(\CM_1) = \int_S \left( 2 c_1(\CL_1) - c_1(\CL_2) + c_1(K_S) \right) \wedge J \ge 0\, ,\\
\label{m2}
m_2 \, \raw \ & A(\CM_2) = \int_S \left( - c_1(\CL_1) + 2  c_1(\CL_2) + c_1(K_S) \right) \wedge J \ge 0\, ,\\
\label{p12}
p_{12} \, \raw \ & A(\P_{12}) = \int_S \left( - c_1(\CL_1) - c_1(\CL_2) + c_1(K_S) \right) \wedge J \ge 0\, .
\end{align}
\end{subequations}
Notice that the product $m_1 m_2 p_{12}$ transforms as a section of $H^0 (S, K_S^3)$ and so, if we would like to consider surfaces of positive curvature, at least one of these three sections should vanish.\footnote{If one is interested in surfaces where the product $m_1 m_2 p_{12}$ can exist, one can see that the D-term equation has a flat direction that allows to reach points in which one of these components vanishes.} Without loss of generality let us take $p_{12} = 0$. Then by using \eqref{dtermsu3} one has that 
\bea\nonumber
\| m_1\|^2 \int_S \left( 2 c_1(\CL_1) - c_1(\CL_2) \right) \wedge J + \|m_2\|^2 \int_S \left( - c_1(\CL_1) + 2  c_1(\CL_2) \right) \wedge J = \\   = - \frac{1}{4\pi} \sum_{i,j} C_{ij} \| m_i\|^2 \| m_j\|^2 < 0
\label{combo3}
\eea
where $C_{ij}$ is the Cartan matrix of $\mg{su}(3)$, and where the last inequality follows from its positive definiteness. Finally, one can derive the set of inequalities 
\be
\int_S c_1(K_S) \wedge J  > \frac{\| m_1\|^2A(\CM_1)+  \| m_2\|^2A(\CM_2)}{\| m_1\|^2+ \| m_2\|^2}  \geq 0 \, ,
\label{finalnogo}
\ee
with the first inequality following from \eqref{combo3} and the second from \eqref{m1} and \eqref{m2}. It is easy to check that a similar result is obtained for any other choice of holomorphic profile for $\Phi$ such that $[\Phi, \Phi^\dagger]$ lies within the Cartan subalgebra. These inequalities are the generalisation of \eqref{ineqnil} for the $SU(2)$ T-brane background and, as in there, they constrain the allowed choices for the bundles $\CL_1$ and $\CL_2$ and forbid a positive curvature for $S$. 

\subsection{Introducing defects}
\label{ss:defects}

Let us now couple a defect theory to the super-Yang-Mills theory on $S$, following \cite{Beasley:2008dc}. Such a theory will be localised on $\IR^{1,3} \times \Sig$, where $\Sig = S \cap S'$ arises from the intersection with a surface $S' \in \CM$ wrapped by a second stack of 7-branes. If the symmetry group of that second stack is $G_{S'}$, then in general there will be matter fields transforming under irreducible representations of $G_S \times G_{S'}$ and localised at the intersection locus $\IR^{1,3} \times \Sig$. The lowest component of such multiplets are complex scalars $(\sig, \sig^c)$ whose internal profile is determined by sections of vector bundles on $\Sig$, namely
\be
\sigma \in \Gamma \left( K_\Sigma^{1/2} \otimes \mc{U} \otimes \mc{U}^{\prime} \right), \qquad \qquad
\sigma^c \in \Gamma \left( K_\Sigma^{1/2} \otimes \mc{U}^* \otimes (\mc{U}^{\prime})^* \right).
\label{sigbun}
\ee 
Here $\mc{U}$, $\mc{U}'$ are the vector bundles associated via the corresponding irreducible representation to the principal bundles on the 7-brane stacks on $S$, $S'$, respectively, and restricted to the curve $\Sig$. Non-trivial vevs for such 4d fields correspond to localised sources for our previous Hitchin system describing the internal 7-brane background. More precisely, from the viewpoint of the 7-brane theory on $S$ we have that the BPS equations \eqref{susy7} are deformed to
\begin{subequations}
\label{susy7d}
\begin{align}
\label{Fterm1d}
\bar \partial_{\mathbb{A}} \Phi = &\, \de_\Sig \wedge  \llangle \sig^c, \sig \rrangle_{\mg{g}_S}\, ,\\
\label{Fterm2d}
\mathbb{F}^{(0,2)} =&\, 0\, ,\\
\label{Dtermd}
J \wedge \mathbb{F} +\frac{1}{2} [\Phi, \Phi^\dagger] =&\, - \oh J \wedge \de_\Sig \, \mu    \, .
\end{align}
\end{subequations}
Here $\de_\Sig$ is the two-form on $S$ with delta-function support along $\Sig$ and which represents the Poincar\'e dual of its homology class. Multiplying it appear the complex outer product $ \llangle \sig^c, \sig \rrangle_{\mg{g}_S}$ and the real moment map $\mu$. Both quantities are bilinear in the defect fields, and in the case that $\mc{U}'$ is a line bundle they locally read\footnote{In the particular case where $\mc{U}$ is split, as will be the case in our discussion below, eqs.\eqref{momc} and \eqref{momr} are in fact globally well-defined.} 
\bea
\label{momc}
\llangle\sigma^c, \sigma\rrangle_{\mg{g}_S} &=& \sigma^c_j (T^I)^j{}_{i} \sigma^{i} \, \mg{t}_I \, , \\
\mu &=& h_\Sigma^{-1/2} \left[h^{\prime \, -1}\bar{\sigma}^{\bar{k}} H_{\bar{k}j} (T^I)^j{}_{i} \sigma^i - h' \sigma^c_i  (T^I)^i{}_{j} H^{j\bar{k}} \bar{\sigma}^c_{\bar{k}} \right] \mg{t}_I \, ,
\label{momr} 
\eea
with $\mg{t}_I$ the generators of $\mg{g}_S = \mr{Lie}(G_S)$ and $T^I$ the representation under which the defect fields $\sig$ transform. In addition $h_\Sigma$ is the hermitian metric on the defect curve, and $H$, $h'$ are the metrics of the bundles ${\cal U}$ and $\mc{U}'$, respectively, that in eq.\eqref{momr} have been restricted to $\Sig$. {We have made these bundle-metric factors explicit in eq. \eqref{momr} for later convenience, because we are going to work in the holomorphic gauge, where they are treated as unknowns of the BPS equations. Consequently, the defect fields appearing in \eqref{momr} are to be regarded as holomorphic sections. On the contrary, in the unitary gauge, the bundle metrics should all be set to the identity, and the defect fields are only covariantly holomorphic, i.e. they satisfy the following equations of motion}
\be
\bar{\del}_{\mb{A}+\mb{A}'} \sigma = \bar{\del}_{\mb{A}+\mb{A}'} \sigma^c = 0 \,,
\label{F1}
\ee
where $\mb{A}$, $\mb{A}'$ act on the appropriate representation and are restricted from $S$, $S'$ to $\Sigma$, respectively. An important point is that, as a consequence of \eqref{sigbun}, $ \llangle \sig^c, \sig \rrangle_{\mg{g}_S}$ can be considered as a $\mg{g}_S$-valued (1,0)-form on $\Sig$, as it is implicitly assumed in \eqref{Fterm1d}. On the other hand $\mu$ is a real scalar. We refer to Appendix \ref{ap:defects} and \cite{Beasley:2008dc} for more details. 

\subsubsection*{A simple setup}

Rather than describing the most general configuration involving defects, let us focus on a simple setup that already shows the new possibilities that adding them brings. Consider type IIB string theory compactified on a  Calabi-Yau three-fold orientifold and a pair of holomorphic four-cycles $S$ and $S'$ within it. The divisor $S$ is wrapped by two D7-branes, therefore hosting a symmetry group $G_S = U(2)$, while $S'$ is wrapped by a single D7-brane and hosts the group $G_{S'} =U(1)$. At their intersection $\Sigma = S \cap S'$, the symmetry group enhances to $G_\Sig = U(3)$, and as a consequence $\Sigma$ localises matter fields $\sig$, $\sig^c$ transforming in the bifundamental representations of $\mg{u}(2) \times \mg{u}(1)$. From the viewpoint of $S$ we will have a 6d defect theory on $\IR^{1,3} \times \Sig$ coupled to the $U(2)$ theory on $\IR^{1,3} \times S$. The existence and internal profile for such defect fields will depend on the worldvolume fluxes threading the four-cycles $S$ an $S'$, restricted to the curve $\Sigma$. By analogy with the setup of section \ref{ss:compact} let us consider a $U(2)$ split gauge bundle $\mathbb{V}=\mathcal{L} \otimes \mc{Q} \oplus\mathcal{L}^{-1} \otimes \mc{Q}$ threading $S$, and a line bundle $\mc{N}$ threading $S'$. It is easy to see that the defect fields are only sensitive to following combination of restricted worldvolume bundles
\be
\hat{\CL}_3 := \CL \big|_\Sigma  \qquad  \qquad  \hat{\CL}_8 := \mc{Q} \big|_\Sigma \otimes \mc{N}^{-1} \big|_\Sigma 
\label{Cartan38d}
\ee
together with the canonical bundle on $\Sig$. In particular we have that \cite{Beasley:2008dc,Blumenhagen:2008zz}
\begin{align}
	\sigma_1 &\in \Gamma \left( K_\Sigma^{1/2} \otimes \hat{\mc{L}}_3 \otimes \hat{\mc{L}}_8 \right), & \sigma_2 &\in \Gamma \left( K_\Sigma^{1/2} \otimes \hat{\mc{L}}_3^{-1} \otimes \hat{\mc{L}}_8 \right)\, , \label{def1} \\
	\sigma_1^c &\in \Gamma \left( K_\Sigma^{1/2} \otimes \hat{\mc{L}}_3^{-1} \otimes \hat{\mc{L}}_8^{-1} \right), & \sigma_2^c &\in \Gamma \left( K_\Sigma^{1/2} \otimes \hat{\mc{L}}_3 \otimes \hat{\mc{L}}_8^{-1} \right). \label{def2}
\end{align}
In terms of the enhancement group $G_\Sig = U(3)$, the bundles \eqref{Cartan38d} can be related to the canonical generators of the $\mg{su}(3)$ Cartan subalgebra or its complexification $\mg{sl}(3)$, see eq.\eqref{ap:Cartan38}. In this sense, one can arrange the different defect fields as entries of the fundamental representation of the $\mg{sl}(3)$ algebra, namely
\be
\left(\begin{array}{ccc}
0 & m & \sig_1 \\ p & 0 & \sig_2 \\ \sig_1^c & \sig_2^c & 0
\end{array}\right)\, .
\label{su3d}
\ee
For completeness, we have also added the modes $m$ and $p$ that extend along the bulk of $S$, and that correspond to elements of $\G (\M)$ and $\G (\P)$, respectively, with $\CM \equiv \mathcal{L}^2\otimes K_S$ and $\P \equiv \mathcal{L}^{-2}\otimes K_S$ defined as above. Notice that there is no field charged under the trace of $\mg{u}(3)$, which completely decouples from the remaining degrees of freedom and will not play any r\^{o}le in the following. In this sense, one may treat this system as a bulk theory with $G_S = U(2)$, coupled to a defect theory with enhanced symmetry group $G_\Sig = SU(3)$.

The complex outer product reads
\bea\nonumber
\llangle\sigma^c, \sigma\rrangle_{\mg{u}(2)} & = &  \sigma_1 \sigma^c_2 \, T_+ + \sigma_2 \sigma_1^c \, T_-  + \oh \left( \sigma_1 \sigma_1^c - \sigma_2 \sigma_2^c \right) \, T_3 + \oh \left( \sigma_1 \sigma_1^c + \sigma_2 \sigma_2^c \right) \, \mf{1}_2 \\
& = & 
\left(
\begin{array}{cc}
\sigma_1 \sigma_1^c &  \sigma_1 \sigma^c_2 \\ 
\sigma_2 \sigma_1^c & \sigma_2 \sigma_2^c \\ 
\end{array}
\right)
\label{outeru2}
\eea
where in the second line we have expressed it in the fundamental representation of $\mg{sl}(2)$. Recall that each of these entries will generate a pole for the Higgs field along the corresponding $\mg{u}(2)$ generator. Having poles along the diagonal entries would correspond to a recombination between the two stacks of D7-branes, and would depart from the $SU(2)$ T-brane profiles of section \ref{ss:compact}. Therefore, in order to reproduce $SU(2)$ T-branes we are left with four possibilities:
\bea
\label{sigh}
\sig_1 = \sig_2 = 0\, , \qquad & &  \qquad \sig_1^c = \sig_2^c = 0\, ,\\
\sig_1 = \sig_2^c = 0\, , \qquad & &  \qquad \sig_1^c = \sig_2 = 0 \, .
\label{sigm}
\eea
Notice that for either choice in \eqref{sigh}, the product \eqref{outeru2} vanishes identically. As a result, in the holomorphic gauge, the Higgs field $\Phi$ needs to have a holomorphic profile, just as in the absence of defects. On the contrary, for either choice in \eqref{sigm} $\Phi$, will develop a pole in one of its off-diagonal entries. As a consequence, $\Phi$ should be described by a meromorphic profile with a pole on top of the defect locus $\Sigma$. In the following we will discuss each of these two possibilities separately, and see how either of them may give rise to an $SU(2)$ T-brane background on $S$, even when $S$ is a four-cycle of positive curvature. 

\subsubsection*{The holomorphic scheme}

Let us first consider the case \eqref{sigh}, with the particular choice $\sig_1^c = \sig_2^c = 0$. The BPS equations on the four-cycle $S$ are given by
\begin{align}
	\pb \,\Phi_S &= 0 \, ,\label{F1h}\\
	\mb{F}_S^{(0,2)} & = 0 \, ,\label{F2h} \\
	J \w \mb{F}_S + \frac{1}{2} [\Phi_S, \Phi_S^\dagger] &= - \frac{1}{2} J \w \delta_\Sigma \, \mu \, . \label{Dh}
\end{align}
where the real moment map expressed in the fundamental representation of $\mg{sl}(2)$ is 
\be
\mu = h_\Sigma^{-1/2} h_8
\left(
\begin{array}{cc}
h_3 |\sig_1|^2 & h_3 \bar{\sig}_1 \sig_2 \\
h_3^{-1} \bar{\sig}_2 \sig_1 & h_3^{-1}  |\sig_2|^2
\end{array}
\right)\, .
\label{muh}
\ee
Here $h_{3} = h_\CL|_\Sig$ and $h_8 = h_\CQ h_\CN^{-1}|_\Sig$ are the metrics for the bundles $\hat{\CL}_{3}$ and $\hat{\CL}_8$ in \eqref{Cartan38d}, defined from the restriction of the metrics $h_\CQ$, $h_\CL$ and $h_\CN$ of the line bundles $\CQ$, $\CL$ and $\CN$, respectively. {As anticipated, we are in the holomorphic gauge, where these metric factors are unknowns of the BPS equations, whereas the defect fields $\sigma_i$ are simply holomorphic sections.} Since we are assuming a split bundle $\mb{V}$ on $S$, the lhs of \eqref{Dh} has vanishing off-diagonal elements, and the same must be true for its rhs. From \eqref{muh} we see that this can be achieved by either setting $\sig_1=0$ or $\sig_2=0$. In manifolds of positive curvature, the appropriate choice is linked to the profile for $\Phi_S$. 

Indeed, let us consider that $\Phi_S$ is given by \eqref{NilpPhi}. Then, if we write $\mb{F}_S = F_0 \mf{1}_2 + F_3 T_3$, the D-term equation \eqref{Dh} amounts to  
\bea \label{DSup}
J \wedge \left(F_0 + F_3\right) & = & - \oh h_\CL^2 \mf{m} \wedge \bar{\mf{m}}\, ,\\
J \wedge \left(F_0 - F_3\right) & = & \oh h_\CL^2 \mf{m} \wedge \bar{\mf{m}} - \frac{h_8}{2 h_3 h_{\Sig}^{1/2}} |\sig_2|^2 J \wedge \de_\Sig\, ,
 \label{DSdown}
\eea
where we have set $\sig_1\equiv0$. On the other hand, the BPS conditions on $S'$ read
\begin{align}
	\pb\, \Phi_{S'} &= 0\ ,\\
	F_{S'}^{(0,2)} & = 0 \ ,\\
	J \w F_{S'} &= \frac{h_8}{2h_3h_\Sigma^{1/2}}  |\sigma_2|^2 J \w \delta_\Sigma\, . 
	\label{DS'}
\end{align}
As an immediate consequence of these equations we have that
\be
2 \int_S J \wedge F_0 + \int_{S'} J \wedge  F_{S'} = 0\, ,
\label{FItrace}
\ee
and so the Fayet-Iliopoulos term for the center-of-mass $U(1)$, the one that corresponds to the trace of $\mg{u}(3)$, vanishes identically. This is consistent with the fact that there is no field charged under this $U(1)$, as pointed out before. In other words, eqs. \eqref{DSup}, \eqref{DSdown} and \eqref{DS'} can be understood as D-term equations for the pair of Cartan generators of $\mg{su}(3)$. In terms of the internal geometry, they can be understood as a Laplace equation on $S$ as follows. We consider the linear combination of the two equations that determines the flux $F_3$, which is given by
\begin{align}
	J \w F_3 = - \frac{1}{2} h_{\mc{L}}^2\, \mf{m} \w \bar{\mf{m}} + \frac{h_8}{4 h_3 h_\Sigma^{1/2}} |\sigma_2|^2 J \w \delta_\Sigma\, . \label{Dlincomb}
\end{align}
Now, similarly to \eqref{decompsu2} we may decompose the flux as
\begin{align}
	F_3 \equiv F_p^{\rm h} + \frac{c}{4} J - i \p \pb g\, ,
\end{align}
where $F_p^{\rm h}$ is primitive and harmonic, $c$ is a constant and $g$ is a global real function. We can now make use of the identity $2i \p \pb g \w J = * \Delta g$, and the {Poincar\'e-Lelong formula} $4\pi i \delta_\Sigma = \mr{d} \mr{d}^c \log |\tau|^2$, with $\tau$ the embedding of $S'$ into $S$, to rewrite \eqref{Dlincomb} as
\begin{align}
\Delta g = c + \frac{h_{\mc{L}}^2}{h_S} |m|^2 - \frac{h_8}{8 \pi h_3 h_\Sigma^{1/2}} |\sigma_2|^2 \Delta \log |\tau|^2 \, ,
\end{align}
where $ |m|^2 = h_S * \mf{m} \wedge \bar{\mf{m}}$. In terms of integrals, their solution is given by
\bea
\label{FI3h}
 \xi_3 = \int_S J \wedge F_3 & = & - \| m \|^2 + \oh \| \sig_2 \|^2\, ,  \\
 \xi_0 = \int_S J \wedge F_0 & = &  - \oh \| \sig_2 \|^2\, ,
\label{FI0h}
\eea
where
\be
\| m \|^2 = \oh \int_S h_\CL^2\, \mf{m} \wedge \bar{\mf{m}}\ , \qquad \qquad \| \sig_2 \|^2 = \oh \int_\Sig \frac{h_8}{2h_3h_\Sigma^{1/2}}  |\sigma_2|^2 J\, .
\label{normh}
\ee
Notice that, whenever $S$ has positive curvature, the existence of the holomorphic section $m$ implies that the lhs of \eqref{FI3h} must be positive. Then, by appropriately tuning the vev of the defect field $\sig_2$, one can find a solution for this system even for this case. Had we chosen instead that $\sig_2 = 0$, the above solution would be replaced by one of the form $ \int_S J \wedge F_3  =  - \| m \|^2 - \oh \| \sig_1 \|^2$ and $\int_S J \wedge F_0  =   - \oh \| \sig_1 \|^2$, and there would be no actual solution in positive curvature manifolds. The r\^oles of $\sig_1$ and $\sig_2$ are reversed if we choose the profile $\Phi_S = p T_-$. Finally, a similar set of solutions can be achieved if in \eqref{sigh} we choose that $\sig_1 = \sig_2 =0$. In particular, $\sig_1^c$, $\sig_2^c$ play the r\^ole of $\sig_2$, $\sig_1$ in the above discussion, respectively.

It is interesting to compare the set of solutions with only $\sig_2 \neq 0$ with the $SU(3)$ T-brane model discussed in section \ref{ss:compact}. Indeed, in the model of section \ref{ss:compact} we turned on the vevs of the pair of fields (${\bf m_1}$, ${\bf m_2}$) and in here we are switching on the pair $(\mf{m}$, $\sig_2$). In terms of the $SU(3)$ symmetry present in both cases these pairs have exactly the same charges, as can be seen from comparing \eqref{FPhisu3} with \eqref{su3d}. From a 4d viewpoint, this implies that these fields have the same charges under the $U(1) \times U(1)$ that survives as a gauge symmetry when worldvolume fluxes are primitive. As a result their D-term potential is the same, as can be seen explicitly by rewriting \eqref{FI3h} and \eqref{FI0h} as
\be
\xi_3 + \xi_0 =  - \| m \|^2 \qquad \qquad 2\xi_0 = - \| \sig_2 \|^2
\label{FIhb}
\ee
which corresponds to \eqref{dtermsu3} with ${\bf p_{12}} \equiv 0$. 

Despite their similarity, in one case  we have a no-go theorem preventing $S$ to have positive curvature, while in the other this obstruction is absent. From the viewpoint of the no-go proof for $SU(3)$, the difference relies on the constraints \eqref{effsu3}, which are modified in the holomorphic scheme. Indeed, while the analogue of \eqref{m1} is still valid in this defect scheme, due to the existence of the bulk mode $m$, eq.\eqref{m2} is dramatically modified. Instead of a positivity condition on $S$, we will have a condition on the degree of the corresponding bundle on the defect curve $\Sig$. Indeed, since $\sig_2$ is a holomorphic section on $\Sigma$, its existence imposes the following condition
\be
{\rm deg}\, \hat{\CL}_8 - {\rm deg}\, \hat{\CL}_3  \geq 1 - g_\Sig \, ,
\label{effdh} 
\ee
where $g_\Sig$ is the genus of $\Sig$.  When going from an $SU(3)$ configuration to the above holomorphic defect scheme, eq.\eqref{m2} is replaced by \eqref{effdh}. Since in general the latter is neither related to the Fayet-Iliopoulos terms \eqref{FIhb} nor to the canonical bundle of $S$, one cannot deduce the first inequality in \eqref{finalnogo}, and the no-go theorem is evaded. 

Of course, in the case where $S'$ is homotopic to $S$, $\Sig$ is a self-intersection curve and \eqref{effdh} and \eqref{m2} can be related. Indeed, one can always see the bundles under which the defect fields are charged as bundles in $S$ restricted to the curve $\Sig$, namely
\be
K_{\Sig}^{1/2} \simeq K_S|_{\Sig}\, , \qquad \qquad \hat{\CL}_3 = \CL_3 |_{\Sig}  \, ,  \qquad \qquad   \hat{\CL}_8=  \CL_8|_{\Sig}  \, ,
\label{extL}
\ee
with ${\CL}_3$, ${\CL}_8$ defined on $S$.
In terms of the bundles $\CL_1$, $\CL_2$ defined below \eqref{FPhisu3} we have that ${\CL}_3 \simeq \CL_1 \otimes \CL_2^{-1/2}$ and ${\CL}_8 \simeq \CL_2^{3/2}$, 
%
%
so considering the bundle
\be
K_S \otimes {\CL}_3^{-1} \otimes {\CL}_8 \simeq K_S \otimes \CL_{-1} \otimes \CL_2^2\,,
\ee
{by continuity} we may assume that $\sig_2$ is the restriction of one of its sections. This promotes the condition \eqref{effdh} to the stronger one \eqref{m2}, and so the inequality \eqref{finalnogo} must be satisfied. We will analyse in greater detail the relation between homotopic four-cycles and defects in self-intersection curves in section \ref{s:Hitchin}.

\subsubsection*{The meromorphic scheme}

Let us now turn to the case \eqref{sigm}, and for concreteness take the choice $\sig_1 = \sig_2^c = 0$. Now the BPS equations on the four-cycle $S$ are given in the holomorphic gauge by
\begin{align}
	\pb\, \Phi_S &=   \de_\Sig \wedge \sigma_2 \sigma_1^c \, T_- \,  ,\label{F1m}\\
	\mb{F}_S^{(0,2)} &= 0 \, ,\label{F2m} \\
	J \w \mb{F}_S + \frac{1}{2} [\Phi_S, \Phi_S^\dagger] &= - \frac{1}{2} J \w \delta_\Sigma \, \mu \, , \label{Dm}
\end{align}
with the real moment map given by
\be
\mu = h_\Sigma^{-1/2} h_3^{-1} 
\left(
\begin{array}{cc}
- h_8^{-1} |\sig_1^c|^2 & 0 \\
0 &  h_8 |\sig_2|^2
\end{array}
\right)\, ,
\label{mum}
\ee
again expressed in the fundamental representation of $\mg{sl}(2)$. Notice that in this case keeping both defect fields with a non-trivial vev is compatible with a split $U(2)$ bundle, and in particular a split $SU(2)$ bundle if we restrict ourselves to $h_8^{-1}|\sig_1^c|^2 =  h_8 |\sig_2|^2$. On the other hand, the non-trivial source term for the Higgs-field F-term suggests that $\Phi_S$ has to be of the form \eqref{NonNilpotentPhiStack}, with at least $\mf{p} \neq 0$. Since we have a split bundle $\mb{V}$, this mode needs to satisfy, in the holomorphic gauge, the F-term equation
\be
\bar{\p} \mf{p}\, =\, \de_\Sig \wedge \sigma_2 \sigma_1^c 
\label{ppole}
\ee
and so it has the profile of a meromorphic section of $\P \equiv \mathcal{L}^{-2}\otimes K_S$. As such, $\P$ does not necessarily need to be effective. Instead, the only requirement that it needs to satisfy is containing meromorphic sections with poles of some order. More precisely, if $\Sig$ is the locus $\{\tau=0\}\subset S$, then we have the identity
\be
\bar{\p} \left(\frac{\p \tau}{\tau^{\,l}}\right) = \frac{2\pi i}{\tau^{\,l-1}} \de_\Sig\,,
\ee
from which one can infer that the pole must be of first order. Therefore, in the absence of holomorphic sections for $\P$, the effectiveness constraint corresponding to the existence of $p$ implies that
\be\label{ExistenceOfPm}
\int_SJ\wedge c_1(\P) + A(\Sig) =\int_SJ\wedge(c_1(K_S)-2c_1(\L)) + A(\Sig) \ge0\, ,
\ee
with $A(\Sig) = \int_\Sig J$. Finally, if the mode $m$ in \eqref{NonNilpotentPhiStack} exists, it must correspond to a holomorphic section, and so the positivity constraint \eqref{ExistenceOfM} applies. Notice that, as before, the product $m p$ transforms as a section of $H^0(S,K_S^2)$, but now the fact that it is meromorphic is not in conflict with $S$ being a manifold of positive curvature.

To build the meromorphic scheme, we will assume that $\P$ only has meromorphic sections, so that \eqref{ExistenceOfPm} applies, and that $\CM$ may contain holomorphic sections, in which case \eqref{ExistenceOfM} would apply. Notice that this implies that both defect fields $\sig_2$ and $\sig_1^c$ have a non-trivial vev. Writing again  $\mb{F}_S = F_0 \mf{1}_2 + F_3 T_3$, the D-terms that correspond to this scenario are
\bea \label{DSupm}
J \wedge F_3 & = & - \oh h_\CL^2 \mf{m} \wedge \bar{\mf{m}}+ \oh h_\CL^{-2} \mf{p} \wedge \bar{\mf{p}}+ \frac{1}{4 h_3h_{\Sig}^{1/2}}  \left( h_8|\sigma_2|^2 + h_8^{-1}|\sig_1^c|^2 \right) J \wedge \de_\Sig \, ,\\
J \wedge F_0 & = & - \frac{1}{4 h_3 h_{\Sig}^{1/2}} \left( h_8 |\sigma_2|^2 - h_8^{-1}|\sig_1^c|^2 \right) J \wedge \de_\Sig\, ,
 \label{DSdownm}
\eea
and the BPS conditions on $S'$ read
\begin{align}
	\pb \,\Phi_{S'} &= 0\ ,\\
	F_{S'}^{(2,0)} & = 0 \ ,\\
	J \w F_{S'} &= \frac{1}{2h_3h_\Sigma^{1/2}} \left( h_8 |\sigma_2|^2 - h_8^{-1} |\sig_1^c|^2 \right) J \w \delta_\Sigma\, .
	\label{DS'm}
\end{align}
As in the holomorphic scheme, we have that the relation \eqref{FItrace} holds. In this case the solution to the two independent D-term equations is given by
\bea
\label{FI3m}
 \xi_3 = \int_S J \wedge F_3 & = & \|p\|^2 - \| m \|^2 + \oh \left( \| \sig_1^c \|^2 + \| \sig_2 \|^2 \right)\ ,  \\
 \xi_0 = \int_S J \wedge F_0 & = &  \oh \left( \| \sig_1^c \|^2 - \| \sig_2 \|^2 \right)\, ,
\label{FI0m}
\eea
with the definitions \eqref{normh} and
\be
\| p \|^2 = \oh \int_S h_\CL^{-2} \mf{p} \wedge \bar{\mf{p}}\ , \qquad \qquad \| \sig_1^c \|^2 = \oh \int_\Sig \frac{ |\sigma_1^c|^2 J}{2h_3 h_8h_\Sigma^{1/2}} \, .
\label{normm}
\ee
Clearly, the simplest set of solutions correspond to those where $\|\sig_1^c \| = \| \sig_2\|$ and $m\equiv0$, so that necessarily $\xi_3 > 0$. Notice that such a FI sign, together with the constraints \eqref{ExistenceOfM} and  \eqref{ExistenceOfPm} imply that
\be
\int_SJ\wedge c_1(\M)> \int_SJ\wedge c_1(K_S)  > - A(\Sig) \, ,
\label{hierarchym}
\ee
which are in principle compatible with manifolds of positive curvature. In general, we expect to find solutions satisfying \eqref{hierarchym} for values of the defect fields $\sig_1^c$, $\sig_2$ and $m$ such that $\xi_3$ is positive and not excessively large. Since the product $\sig_1^c \sig_2$ sources the meromorphic profile for $p$, one would presume that its contribution to the D-term is fixed by their value. The analysis of sections \ref{s:4d} and \ref{s:Hitchin} will provide a more precise picture to this expectation. Finally, when compared to the $SU(3)$ T-brane model discussed in section \ref{ss:compact}, we get a very similar set of D-term equations
\be
\xi_3 + \xi_0 =  - \| m \|^2 + \| \sig_1^c \|^2 + \|p\|^2\,, \qquad \qquad 2\xi_0 = - \| \sig_2 \|^2 + \| \sig_1^c \|^2 \,,
\label{FImb}
\ee
which is essentially \eqref{dtermsu3} with the dictionary $({\bf m_1}, {\bf m_2},{\bf p_{12}}) \raw (m, \sig_2, \sig_1^c)$ and the addition of the contribution from $p$. One can check that adding a contribution of this form to \eqref{dtermsu3} would not change the results below, in the sense that \eqref{finalnogo} would still be valid and positive curvature manifolds excluded. Again, the fact that we may construct T-brane backgrounds with $S$ of positive curvature using the meromorphic scheme is due to the different effectiveness constraints imposed by this class of constructions. We have that these imply eqs.\eqref{ExistenceOfM}, \eqref{ExistenceOfPm} and the conditions derived from the fact that $\sig_1^c$ and $\sig_2$ are holomorphic sections on $\Sig$:
\be
{\rm deg}\, \hat{\CL}_3   \leq  g_\Sig -1 \, , \qquad \qquad {\rm deg}\, \hat{\CL}_8 \leq  g_\Sig -1 - {\rm deg}\, \hat{\CL}_3 \, .
\label{effdm} 
\ee
As before, these conditions are unrelated to the values of $\xi_3$ and $\xi_8$, except in some specific cases like when $\Sig$ is the self-intersection curve of $S$, and we take $\sig_2$ and $\sig_1^c$ to be the restriction of holomorphic sections of the corresponding bundles on $S$. 


\section{The 4d perspective}
\label{s:4d}

In this section we would like to take a four-dimensional perspective on the meromorphic scheme introduced in section \ref{s:defects}. In particular, we will analyse the space of F-flat directions around the point in moduli space in which $\langle \Phi \rangle=0$, 
and deduce how the vevs of the various 4d fields are constrained. As a warm-up, we first work in the absence of defects, where as expected we will find that the massive KK modes of all fields must all have vanishing vev. Then we will add the defect contribution to the superpotential \cite{Beasley:2008dc} and look for solutions to the F-term equations with two of the intersection fields having non-vanishing vev. As in section \ref{s:defects} we will focus, for definiteness, on the fields $\sigma_1^c$ and $\sigma_2$, i.e. those responsible for creating a pole for $\Phi$ as in \eqref{F1m}. In this case, we find two important results. On the one hand, from integrating such an F-term equation, truncated at the zero-mode level, we get a necessary condition for solving it. On the other hand, expanding in a basis of KK modes, we realise that the singular profile for $\Phi$ can be understood as a sum of non-trivial vevs for massive KK replicas, rather than that of a zero mode.\footnote{This in turn is analogous to what happens to the KK modes of the Cartan vector field in the presence of a non-primitive flux (see Appendix B of \cite{Marchesano:2017kke}).} 

Schematically, upon dimensional reduction on a four-cycle $S$ of positive curvature (or simply without holomorphic deformations) we find a superpotential of the form
\be
W = \sum_\a \mu_\a \Psi_\a \Phi_\a -  c_\a \Psi_\a \sig_1^c \sig_2\,,
\label{Wsketch}
\ee
where $\Psi_\a$ and $\Phi_a$ run over the KK modes of (0,1) and (2,0)-forms, respectively, of mass $\mu_\a$ and $\sig_1^c$, $\sig_2$ are the 4d fields corresponding to the massless defect modes. As a vev is given to the defect fields $\sig_1^c$, $\sig_2$ the F-term for $\Phi_\a$ implies that
 \be
\langle \Phi_\a \rangle  = \frac{c_\a}{\mu_\a} \langle \sig_1^c \sig_2 \rangle\,,
\ee
and so the massive $(2,0)$-forms develop a vev due to their coupling to defects. When such a vev is combined with their wavefunction along $S$, one obtains a profile that reproduces the pole of the meromorphic scheme. Finally, the couplings of the form $c_0 \Psi_0 \sig_1^c \sig_2$, with $\Psi_0$ a zero mode, provide an obstruction to giving a vev to the product $\sig_1^c \sig_2$ and so to realising the meromorphic scheme. When $S$ has holomorphic deformations, extra Yukawa coupling involving $\Psi_0$ must be added to \eqref{Wsketch}, modifying the corresponding F-term for $\Psi_0$ and relaxing the above obstructions. In the following we will sketch the main idea of this computation,  deferring all technical details of this discussion to Appendix \ref{ap:4d}, where we also give a complete presentation of the F-term constraints.

\subsubsection*{The spectrum of bulk KK modes}

We start with some preliminary material which will allow us to perform the dimensional reduction from 8d to 4d. The Hodge duality operation can be defined to act as follows on the space of $(p,q)$-forms of the internal K\"ahler surface:
\be
*:\Omega^{(p,q)}\longrightarrow\Omega^{(2-q,2-p)}\,.
\ee
This allows us to define the adjoint of the Dolbeault operator
\be\label{adjointd}
\bar{\p}^{\dagger}=-*\p*
\ee
with respect to the hermitian product on $S$
\be
\int_S\alpha\wedge*\bar{\beta}\,,
\ee
for any two $(p,q)$-forms $\alpha,\beta$. In these conventions the $(2,0)$-forms, which are all primitive, are self-dual \cite{Ballmann}. Hence, the holomorphic entries of the Higgs field $\Phi$ are all harmonic, self-dual forms. For the purpose of this section, however, we will need to take into account the non-zero modes of $\Phi$ too, and thus the space of exact $(2,0)$-forms. Let $\{\chi^A\}_A$ be a complete basis for the space of $(2,0)$-forms, normalised such that
\be
\frac{1}{V_S}\int_S\chi^A\wedge *\bar{\chi}^B=\frac{1}{V_S}\int_S\chi^A\wedge \bar{\chi}^B=\de^{AB}\,,
\label{ortochi}
\ee
where, in the first step, we applied self-duality. We take each of the elements of this basis to be eigenstates of the Laplacian operator $\Delta\equiv \bar{\p} \bar{\p}^\dagger +\bar{\p}^\dagger \bar{\p}$. For future convenience, let us split the collective index $A$ as $(\alpha_0,\alpha)$, to divide the basis into zero and non-zero modes. That is, we have that $\Delta\chi_{\alpha_0}=0$ and $\Delta\chi_{\alpha}=-k^2_\alpha\chi_\alpha$. In order to expand the KK modes of the vector field, we also pick a complete basis of $(0,1)$-forms, $\{\psi^I\}_I$ normalised such that
\be
\frac{1}{V_S}\int_S\psi^I\wedge *\bar{\psi}^J=\de^{IJ}\,.
\label{ortopsi}
\ee
As before, we take them to be eigenstates of the Laplacian, and separate the zero-modes indicated with the index $i_0$ from the non-zero modes indexed by $i$, so that $\Delta\psi^{i_0}=0$ and $\Delta\psi^i=-l^2_i\psi^i$. It turns out that the subspace of \emph{exact} $(2,0)$-forms is isomorphic to the subspace of \emph{coexact} $(0,1)$-forms. They are mapped into each other by a pair of invertible matrices $\mu,\tilde{\mu}$ as follows
\bea
\bar{\p}{\psi^i}&=&i\mu^i_\alpha\bar{\chi}^\alpha\,,\\
\bar{\p}^\dagger\bar{\chi}^\alpha&=&i\tilde{\mu}^\alpha_i\psi^i\,.\label{eqforchi}
\eea
By applying $\Delta$ to any of the above equations, and using that $[\Delta,\bar{\p}]=[\Delta,\bar{\p}^\dagger]=0$, one easily finds that the eigenvalues $k^2_\alpha$ and $l^2_i$ must be equal to each other, and in this sense the indices $\a$ and $i$ can be identified. Moreover, by applying $\bar{\p}^\dagger$ to the first equation and $\bar{\p}$ to the second one, we get the following set of equations respectively:
\bea\label{masseq}
\mu^i_\alpha\tilde{\mu}^\alpha_j&=&l^2_j\de^i_j\,,\nonumber \\
 \tilde\mu^\alpha_i\mu^i_\beta&=&k^2_\beta\de^\alpha_\beta\,,
\eea
with no sum on the rhs. This gives $\mu,\tilde{\mu}$ the meaning of (complex) mass matrices.

Note that all the $(p,q)$-forms we will deal with are bundle-valued. Since we consider a split $SU(2)$ bundle, this amounts to having three different basis of $(2,0)$-forms $\{\chi_{\bullet A}\}_A$ and three different basis of $(0,1)$-forms $\{\psi_\bullet^I\}_I$, where $\bullet=\{+,-,3\}$ runs over the generators of $\mg{sl}(2)$ (cf. Appendix \ref{ap:su3}) and indicates positive, negative and zero charge respectively. Each of these basis will satisfy  orthonormality relations of the form \eqref{ortochi} and \eqref{ortopsi}. Accordingly, we will have to consider three different pairs of mass matrices in \eqref{masseq}.

\subsubsection*{F-terms without defects}

Let us start with the case where no defects are present. As is well known \cite{Donagi:2008ca,Beasley:2008dc}, the holomorphic sector is regulated by the following 4d superpotential
\be
W_S=\int_S{\rm Tr}\;\mathbb{F}\wedge\Phi\,,
\label{WS}
\ee
which imposes that $\bar{\p}_{\mathbb{A}}\Phi=\mathbb{F}^{(0,2)}=0$. We are now interested in studying the space of infinitesimal fluctuations for an $SU(2)$ Hitchin system, around a BPS background such that $\langle\Phi\rangle=0$, $\langle\mathbb{A}\rangle=AT_3$ and $\mathbb{F}\wedge J=0$. By working in the holomorphic gauge, we can simply ignore the vacuum profile for $A$, and hence have
\bea\label{pPhi}
\bar{\p}_{\mathbb{A}}\Phi&=&\bar{\p}\,\delta\Phi +[\delta\mathbb{A}^{(0,1)},\delta\Phi]= 0 \, , \\
\mathbb{F}^{(0,2)}&=&\bar{\p}\,\de\mathbb{A}^{(0,1)}-\frac{i}{2}[\delta\mathbb{A}^{(0,1)},\delta\mathbb{A}^{(0,1)}]=0\,,\label{F02}
\eea
where we have defined the matrices of fluctuations
\be
\delta\Phi=\left(\begin{array}{cc}\mf{v}&\mf{m}\\ \mf{p}&-\mf{v}\end{array}\right)\,,\qquad\delta\mathbb{A}^{(0,1)}=\left(\begin{array}{cc}a_3 &a_+\\ a_-& -a_3\end{array}\right)\,.
\label{fluctu}
\ee
It is immediate to see that, from the off-diagonal components of \eqref{pPhi} and from the diagonal component of \eqref{F02} we get respectively the three F-term equations
\bea
\bar{\p}\mf{m}&=&2a_+\wedge \mf{v}  - 2a_3 \wedge \mf{m}\, , \label{a-}\\
\bar{\p}\mf{p}&=&-2a_-\wedge \mf{v} +  2 a_3 \wedge \mf{p}\,, \label{a+}\\
\bar{\p} a_3 &=&i a_+\wedge a_-\label{v}\,.
\eea

One can see that these and the remaining F-terms constraints imply a vanishing vev for the massive KK modes, as discussed in Appendix \ref{ap:4d}. 
 At the massless level, the modes for $a_3$ are absent if, as in previous sections, one assumes that $S$ is simply connected. Moreover, at the massless level the lhs of eqs.\eqref{a-}-\eqref{v} vanish, and the rhs gives constraints on the vev of the zero modes. These can be parametrised, e.g.,  by wedging the rhs of \eqref{a-} with an arbitrary zero mode in the $a_-$ sector, namely $\hat{a}_- = a_{-\, i_0} \psi_-^{i_0}$, and integrating over $S$. Similarly, one may wedge the rhs of \eqref{a+} with the zero modes $\hat{a}_+ = a_{+\, j_0} \psi_+^{j_0}$ and the rhs \eqref{v} with $\hat{\mf{v}}=v_{\a_0} \chi_{3}^{\a_0}$, and integrate over $S$. One ends up with the  constraints
\be\label{FtermsNoDefects}
\Lambda^{i_0j_0\a_0}_{3} a_{+\, j_0} v_{\a_0}=\Lambda^{i_0j_0 \a_0}_{3} a_{-\, j_0} v_{\a_0}=\Lambda^{i_0j_0\alpha_0}_{3} a_{+\, i_0} a_{-\, j_0}=0\,,
\ee
where the indices $i_0$, $j_0$ and $\a_0$ run over the subspace of zero-modes, and we have defined the Yukawa couplings
\be
\Lambda^{IJA}_{3}=\int_S \psi_+^I\wedge\psi_-^J\wedge\chi_{3}^{A}\,.
\ee
Alternatively, one may derive \eqref{FtermsNoDefects} from the superpotential truncated at zero mode level.
\be
W\, =\, \Lambda^{i_0j_0\a_0}_{3} a_{+\, j_0} a_{-\, j_0} v_{\a_0}\,.
\label{yukzero}
\ee
From equations \eqref{FtermsNoDefects} it is clear that at least two among the three sets of massless modes $\{a_{+\, i_0}\}_{i_0}$, $\{a_{-\, i_0}\}_{i_0}$ and $\{v_{\a_0}\}_{\a_0}$ must attain trivial vacuum expectation values.

\subsubsection*{F-terms with defects}

Let us now introduce defects and see how equations \eqref{FtermsNoDefects} are modified, inducing non-trivial vevs for non-zero KK modes. As anticipated, this will be the 4d counterpart of the meromorphic profile introduced in section \ref{s:defects}.

Defects are localised on the curve $\Sigma\subset S$ and affect the holomorphic sector through the superpotential \cite{Beasley:2008dc}
\be
W_\Sigma=\int_S \de_\Sigma\wedge\llangle \sigma^c, \bar{\p}_{\mathbb{A}}\sigma \rrangle_{\mathfrak{g}_S}\,.
\label{Wdef}
\ee
For definiteness, let us consider non-trivial vevs for the defect fields $\sigma_1^c$ and $\sigma_2$, which, as seen in section \ref{s:defects}, generate a first-order pole for the Higgs field along $\Phi = \mf{p} T_-$, see eq.\eqref{ppole}. Hence, while equations \eqref{a-} and \eqref{v} remain unmodified, equation \eqref{a+} becomes
\be\label{newa+}
\bar{\p}\mf{p}=-2a_-\wedge \mf{v} +  2 a_3 \wedge \mf{p} +\de_\Sigma\wedge \sigma_1^c\sigma_2\,.
\ee
The fact that eqs.\eqref{a-} and \eqref{v} remain intact implies, by the discussion of Appendix  \ref{ap:4d}, that all the massive modes of the sectors $a_3$, $a_\pm$ must have vanishing vevs, and that the same must be true for the sectors $\mf{m}$ and $\mf{v}$. The sector $\mf{p}$ is however more subtle due to the defect contribution on \eqref{newa+}. Indeed, let us decompose an arbitrary profile in this sector as a sum of KK modes, as $\mf{p} = p_A \chi_{-}^{A}$. On the one hand, since the modes $a_3$ have all zero vev, the potential zero mode contribution $p_{\a_0} \chi_{-}^{\a_0}$ will not play any role in eq.\eqref{newa+}. On the other hand, the massive modes satisfy  the following relation
\be\label{neweqforchi}
\bar{\p}\chi_{-}^{\alpha}=i(\bar{\tilde{\mu}}_{-})^\alpha_i *\bar{\psi}_{-}^{i}\, ,
\ee 
where we have used the self-duality of $\chi_{-}^{\alpha}$ and the definition \eqref{adjointd} to rewrite the complex conjugate of \eqref{eqforchi}. So plugging the massive mode expansion $p_\a  \chi_{-}^{\a}$ into \eqref{newa+} one obtains
\be\label{finalEq}
p_\alpha = -\frac{i}{k_{-\, \alpha}^2}\bar{\mu}^\alpha_{-\, i}\left(\Gamma^{i \kappa_1 \kappa_2}_{+}\langle \sigma_1^c\rangle_{\kappa_1} \langle\sigma_2\rangle_{\kappa_2} - 2 \Lambda^{ij_0 \beta_0}_{3} a_{-\, j_0} v_{\beta_0}\right)\, ,
\ee
where we have defined
\be
\Gamma^{I \kappa_1 \kappa_2}_{+} = \int_S \de_\Sigma\wedge (\sigma_1^c)^{\kappa_1} (\sigma_2)^{\kappa_2} \wedge \psi^I_+\, ,
\ee
with $\kappa_1$, $\kappa_2$ running over the defect zero modes of the kind $\sigma_1^c$, $\sigma_2$, respectively. Recall that the indices $\alpha,i$ run over the subspace of \emph{non-zero} modes, such that we could invert the mass matrix and make use of \eqref{masseq}. 

The F-term condition can be derived from a superpotential of the form
\be
W =  \mu_\a \Psi_\a \Phi_\a -  \Gamma^{\a \kappa_1 \kappa_2} \Psi_\a (\sig_1^c)_{\kappa_1} (\sig_2)_{\kappa_2} + \Lam^{\a j_0 \b_0} \Psi_\a a_{-\, j_0 } v_{\b_0}\, ,
\label{Wforp}
\ee
with $\Psi$ and $\Phi$ representing the massive modes of the $a_+$ and $\mf{p}$ sector, respectively. If we assume that the surface deformation modes $v_{\b_0}$ are absent, one essentially recovers the superpotential \eqref{Wsketch}, and one can understand the meromorphic profile for $\mf{p}$ as a sum of massive modes with non-trivial vev, sourced by the vevs of the defect modes. 

To extract the implications of \eqref{newa+} at the zero mode level, let us wedge both sides of this equation by a zero mode in the $a_-$ sector, $\hat{a}_- = a_{-\, i_0} \psi_-^{i_0}$, and then integrate over $S$. Since in the holomorphic gauge each of the elements of this basis has a holomorphic profile, the left-hand-side is a total derivative and its integral over $S$ vanishes. From the integral on the rhs we then get the equations
\be\label{FernandoConstraint}
\Gamma_+^{i_0 \kappa_1 \kappa_2}\langle \sigma_1^c\rangle_{\kappa_1} \langle\sigma_2\rangle_{\kappa_2} - 2 \Lambda^{i_0j_0\alpha_0}_{3} a_{-\, j_0} v_{\alpha_0} = 0\, .
\ee
Notice that again \eqref{FernandoConstraint} can be deduced from a superpotential of the form \eqref{Wforp}, upon extending the index $\a\raw A$ to  include the zero modes of the corresponding sector $\Psi$, the term \eqref{yukzero} and the coupling of defects to $\hat{a}_+$. Alternatively, one could have obtained such superpotential directly from \eqref{WS} and \eqref{Wdef}, by plugging the expansion on KK modes for each of the sectors in \eqref{fluctu} and truncating the appropriate towers of massive modes. 

The 4d F-term constraints \eqref{FernandoConstraint} are necessary conditions for the existence of solutions of \eqref{newa+}, and play an analogous r\^ole of the 4d D-terms equations obtained in section \ref{s:defects}, see e.g. eqs.\eqref{FI3h} and \eqref{FI0h}. They are of course trivial in the absence of zero modes in the $a_-$ sector. However, if such modes are present, eq.\eqref{FernandoConstraint} may represent an obstruction to constructing the meromorphic scheme. Such an obstruction is particularly relevant for surfaces without complex deformations, like e.g. surfaces of positive curvature, for which there is no zero mode $v_{\a_0}$. For instance, in the case in which there is a single defect mode of the kind $\sig_1^c$ and of the kind $\sig_2$, \eqref{FernandoConstraint} would be impossible to satisfy unless the topological coupling $\Gamma_+^{i_0 \kappa_1 \kappa_2}$ vanishes. When several defect modes are present, one should find a combination such that the sum over $\kappa_1$, $\kappa_2$ in \eqref{FernandoConstraint} vanishes for all choices of $i_0$, while the ones in \eqref{finalEq} do not. Finally, notice that when the zero modes $v_{\a_0}$ exist and we manage to solve \eqref{FernandoConstraint} by giving them a vev, we are strictly speaking no longer in the T-brane scenario of section \ref{ss:defects}, since now the $\mg{su}(2)$ sector corresponds to two intersecting 7-branes rather than to two coincident ones.


\section{Defects and Hitchin systems}
\label{s:Hitchin}

As described in section \ref{s:defects}, defect fields arise at the intersection of two stacks of 7-branes wrapping holomorphic four-cycles $S$ and $S'$. One particular case is when $S$ has an effective canonical bundle $K_S$, and $S'$ is a holomorphic deformation of $S$. The curve $\Sigma$ hosting the defects is, in this case, the self-intersection curve of $S$, which represents the Poincar\'e dual of $c_1(K_S)$. Interestingly, the enhancement group $G_\Sig \supset G_S \times G_{S'}$ over this curve can now be extended to the whole of $S$, in the sense that this is the symmetry group of the system when $S$ and $S'$ coincide. Therefore the information of the whole system, including the defects, should be contained in a regular Hitchin system with group $G_\Sig$, and the BPS equations with defects discussed above should be recovered in the limit in which the intersection fields are ultra-localised at $\Sigma$.

In the following we would like to explore the dictionary between Hitchin systems on self-intersecting curves and systems with defects in further detail, in order to understand how to recover the latter from the former. To simplify our discussion we will consider a setup with enhanced gauge group $G_\Sig = SU(3)$, in which two D7-branes wrap $S$ and a third one is its holomorphic deformation $S'$. This will allow to easily connect with the simple defect setup analysed in the previous section, and in particular with the holomorphic and meromorphic schemes discussed there. As we will see, from the viewpoint of the $SU(3)$ Hitchin system these two configurations are not that different.

\subsection{The meromorphic scheme}

Let us consider a  Hitchin system with gauge group $SU(3)$, defined on a surface $S$ with effective canonical bundle $K_S$. We introduce a Higgs field which in the holomorphic gauge reads
\be
\Phi^{\rm h} \, =\,  \frac{1}{3} 
\left(
\begin{array}{ccc}
{\bf v} & 0 & 0 \\
0 & {\bf v} & 0 \\
0 & 0 & -2{\bf v}
\end{array}
\right)
+
\left(
\begin{array}{ccc}
0 & 0 & 0 \\
 0 & 0 & {\bf s}_2 \\
{\bf s}_1^c & 0 & 0
\end{array}
\right)\, .
\label{hPhim}
\ee
On the one hand, ${\bf v} \in H^{(2,0)}(S)$ corresponds to a holomorphic deformation of the cycle $S$. The piece of $\Phi$ proportional to {\bf v} has the effect of separating the initial $SU(3)$ stack into two stacks of two and one 7-branes, each wrapping surfaces homotopic to $S$ and intersecting at the curve $\Sig = \{ {\bf v} = 0\} \subset S$. On the other hand, the (2,0)-forms ${\bf s}_2$,  ${\bf s}_1^c$ can be considered to be sections of line bundles on $S$. Indeed, notice that in the limit of coincident 7-branes ${\bf v} \raw 0$ the system reduces to the $SU(3)$ system in \eqref{FPhisu3}, upon the identifications $\mf{m}_2 \leftrightarrow \mf{s}_2$, $\mf{p}_{12} \leftrightarrow \mf{s}_1^c$. There we may consider a split gauge bundle with a corresponding worldvolume flux of the form
\be
\mathbb{F} = F_3\, H_3 + F_8\, H_8 =
\left(\begin{array}{ccc}
F_3+\frac{1}{3}F_8&0 &0 \\ 0 & - F_3+ \frac{1}{3}F_8& 0 \\ 0 & 0 & -\frac{2}{3}F_8
\end{array}\right)\, , 
\label{Fsu3m}
\ee
where the $\mg{su}(3)$ Cartan generators $H_3$, $H_8$ are defined in terms of the canonically normalised ones in \eqref{ap:Cartan38} as $H_3 = 2H_1$ and $H_8 = \frac{2}{\sqrt{3}}H_2$. As usual, the $(1,1)$-forms $F_i$, $i=3,8$ are related to the corresponding line bundles as $[F_i] = 2\pi c_1(\CL_i)$. The particular choice of flux in \eqref{Fsu3m} allows to relate the corresponding bundles with the pair $\hat{\CL}_3$, $\hat{\CL}_8$ that appear in the defect schemes of subsection \ref{ss:defects}, or more precisely to identify them with the extensions $\CL_3$, $\CL_8$ introduced around eq.\eqref{extL}. Using the relation specified there with the bundles $\CL_1$, $\CL_2$ that correspond to the flux in \eqref{FPhisu3} one finds that 
\be
\mf{s}_2 \in H^{2,0} \left({\CL}_3^{-1} \otimes {\CL}_8\right)\, , \qquad \qquad \mf{s}_1^c \in H^{2,0} \left({\CL}_3^{-1} \otimes {\CL}_8^{-1}\right)\, .
\label{scharges}
\ee
As we turn on the four-cycle deformation $\mf{v}$, the flux \eqref{Fsu3m} will no longer yield a solution to the D-term equation \eqref{Dterm}, and we will need to consider a non-split bundle. In general, for non-split bundles one may not identify individual entries of $\Phi$ as sections of line bundles as done above. However, as in our case the no-split bundle is continuously connected to a split one in the limit $\mf{v} \raw 0$, one may impose \eqref{scharges} for arbitrary values of $\mf{v}$.\footnote{Mathematically, our gauge bundle is a holomorphic vector bundle of rank $2$, whose structure group is reduced from $SU(3)$ to the Cartan torus $U(1)\times U(1)$. However, due to the ${\bf v}$-mode in $\Phi^{\rm h}$, supersymmetry requires it to carry a connection (or equivalently a hermitian metric) whose holonomy is \emph{not} in $U(1)\times U(1)$. Nevertheless, since the topology of the bundle is such that it admits a split form, the group of physical gauge transformations a solution of the Hitchin system is defined up to is still $U(1)\times U(1)$, and thus the various entries of $\Phi^{\rm h}$ are still globally well-defined as sections of line-bundles, like in eq. \eqref{scharges}.}

The information of the non-split bundle is encoded in the complexified gauge transformation that allows to solve the D-term equations. Let us take it to be of the form
\be
B \, =\,
\left(
\begin{array}{ccc}
e^{f_3/2+f_8/6} & 0 & 0 \\
0  & e^{-f_3/2+f_8/6} & 0  \\
0 & 0 & e^{-f_8/3}
\end{array}
\right)
\cdot
\left(
\begin{array}{ccc}
1 & 0 & 0 \\
- \oh \xi_1^c \xi_2  & 1 & - \xi_2  \\
 \xi_1^c & 0 & 1
\end{array}
\right)
\label{Bxi}
\ee
with $\xi_1^c \in \G(\CL_3^{-1} \otimes \CL_8^{-1})$ and $\xi_2 \in \G(\CL_3^{-1} \otimes \CL_8)$.  The unitary-gauge Higgs field then is
\be
\Phi = B \Phi^{\rm h} B^{-1} = 
  \frac{1}{3} \mf{v} \Id_3 +
\left(
\begin{array}{ccc}
0 & 0 & 0 \\
- e^{-f_3}\left(\mf{s}_1^c \xi_2 + \xi_1^c \mf{s}_2 + \mf{v}\xi_1^c\xi_2\right) & 0 & e^{-\frac{f_3}{2}+\frac{f_8}{2}} \left( \mf{s}_2 + \mf{v} \xi_2\right)  \\
e^{-\frac{f_3}{2} -\frac{f_8}{2}} \left( \mf{s}_1^c + \mf{v} \xi_1^c \right) & 0 & -\mf{v}
\end{array}
\right)\, ,
\label{uPhixim}
\ee
whose individual entries are globally well-defined (2,0)-forms in $S$. One the one hand, one expects that the sections $\xi_1^c$, $\xi_2$ vanish in the limit $\mf{v} \raw 0$. On the other hand, as we increase the vev of the deformation $\mf{v}$, they should implement the localisation of the unitary profile for the fields $\mf{s}_1^c$, $\mf{s}_2$ along $\Sig$. We find that an appropriate choice to reproduce both features is 
\be
\xi_1^c \, =\, \frac{s_1^c}{v}  \, \left(e^{-\lam |v|^2}-1 \right)\, , \qquad \qquad \xi_2 \, =\, \frac{s_2}{v}  \, \left(e^{-\lam |v|^2 }-1 \right)\, ,
\label{xis}
\ee
where $v$, $s_1^c$, $s_2$ are the scalar holomorphic sections that correspond to $\mf{v}$, $\mf{s}_1^c$, $\mf{s}_2$. In addition, $\lam$ is of the form
\be
\lam = \frac{\lam_\star}{\sqrt{|g_S|}}
\label{deflam}
\ee
with $\lam_\star$ a globally well-defined real function on $S$ that, for most purposes of the discussion below, can be considered to be a constant. Notice that away from the self-intersection locus $\Sig = \{ v=0\}$, the exponential factor in \eqref{xis} can be neglected, and $\xi_1^c$, $\xi_2$ become the entries that take \eqref{hPhim} into its Jordan canonical form. Near $\Sig$ the exponential becomes relevant and renders $\xi_1^c$, $\xi_2$ regular. In fact, they both vanish at $v=0$, so their effect on $\Phi^{\rm h}$ will be irrelevant near this locus. Indeed, plugging \eqref{xis} into \eqref{uPhixim} one obtains
\be
\Phi \, =\,  \frac{1}{3}\mf{v} \Id_3 +
\left(
\begin{array}{ccc}
0 & 0 & 0 \\
e^{-f_3}  \frac{s_1^c s_2}{v^2}  \left(1- e^{-2\lam  |v|^2} \right) \mf{v} & 0 &  e^{-\frac{f_3}{2} + \frac{f_8}{2}}  e^{- \lam  |v|^2} \mf{s}_2 \\
e^{-\frac{f_3}{2} - \frac{f_8}{2}}  e^{- \lam  |v|^2} \mf{s}_1^c & 0 & -\mf{v}
\end{array}
\right)\, ,
\label{uPhim}
\ee
which displays a clear localisation of the fields $s_1^c$, $s_2$ around the self-intersection locus via the exponential factor $e^{- \lam |v|^2}$. {In fact, the entries of \eqref{uPhim} linear in $\mf{s}_1^c$ or $\mf{s}_2$ correspond to the wavefunction profile along the Higgs-field component that one obtains for the zero mode fluctuations at the intersection of two 7-branes}, cf. \cite{Font:2009gq,Cecotti:2009zf,Conlon:2009qq,Aparicio:2011jx,Font:2012wq}. The remaining off-diagonal entry is also localised around $\Sigma$ but unexpected from the viewpoint of such a wavefunction analysis, which only detects up to a linear dependence on intersection fields. We will however see below that it corresponds to the appearance of the pole of the meromorphic defect scheme. 

In this unitary gauge the $\mg{su}(3)$ gauge connection is given by
\bea\nonumber
i \mb{A}^{(0,1)} \, =\, - B \, \bar{\p}\, B^{-1} & = & 
\left(
\begin{array}{ccc}
\oh  \bar\p f_3 + \frac{1}{6} \bar\p  f_8& 0 & 0 \\
 0 &  - \oh \bar\p f_3 + \frac{1}{6} \bar\p  f_8  & 0 \\
0 & 0 & \frac{1}{3} \bar\p  f_8
\end{array}
\right) \\
& + &
\left(
\begin{array}{ccc}
 0 & 0 & 0 \\
 e^{-f_3}  \oh \left(\xi_1^c \bar{\p} \xi_2 - \xi_2 \bar{\p} \xi_1^c\right) & 0 &  e^{-\frac{f_3}{2} + \frac{f_8}{2}} \bar{\p} \xi_2 \\
 - e^{-\frac{f_3}{2} - \frac{f_8}{2}} \bar{\p} \xi_1^c & 0 & 0
\end{array}
\right)   \, ,
\label{uAxim}
\eea
which after plugging the Ansatz \eqref{xis} becomes
\be
i \mb{A}^{(0,1)}  = \oh H_3  \bar\p f_3 + \oh H_8 \bar\p  f_8 + 
\left(
\begin{array}{ccc}
 0 & 0 & 0 \\
 0 & 0 &  - e^{-\frac{f_3}{2} + \frac{f_8}{2}} s_2 \\
 e^{-\frac{f_3}{2} - \frac{f_8}{2}} s_1^c & 0 & 0
\end{array}
\right)  e^{-\lam |v|^2} \bar{\p} \left( \lam \bar{v}\right)  \, .
\label{uAm}
\ee
Notice that as expected the $\mg{su}(3)$ bundle does not have a split connection but, due to this particular Ansatz, we recover a split one if we restrict ourselves to the $\mg{u}(2)$ subalgebra that contains $H_3$. This is in agreement with the simple defect setup discussed in section \ref{ss:defects}, and in particular with the meromorphic scheme that we are trying to reproduce. In addition, note that the off-diagonal entries in \eqref{uAm} reproduce the expected wavefunction profile along the gauge boson components for the fluctuations of fields localised at 7-brane intersections. 

By construction, the profiles for $\Phi$ and $\mb{A}$ in the unitary gauge satisfy the F-term equations of the Hitchin system. The next step is then to introduce such profiles into the D-term equation \eqref{Dterm} to find a solution in terms of $f_3$, $f_8$ and $\lam$. For simplicity, let us consider the particular case where $s_1^c$ and $s_2$ differ by a constant phase, so that we can rewrite the D-term equations in terms of a section $s$ such that  $s = e^{-i\varphi_1} s_1^c  = e^{-i\varphi_2} s_2$. This can only be a solution if the bundle $\CL_8$ is trivial, so we may take $f_8 =0$. Then, one can see that the following structure is recovered
\be
J \wedge \mathbb{F} +\frac{1}{2} [\Phi, \Phi^\dagger] = 
\left(
\begin{array}{ccc}
 C & 0 & e^{-i\varphi_1} \bar{D} \\
 0 & - C &  e^{i\varphi_2} D \\
 e^{i\varphi_1} {D} & e^{-i\varphi_2} \bar{D} & 0
\end{array}
\right)
\label{Dtermmero}
\ee
and the D-term equation reduces to two independent differential equations $C=D=0$ with unknowns $f_3$ and $\lam$. The off-diagonal components of \eqref{Dtermmero} vanish if one imposes
\be
2 i J \w  \p \left( s\,  e^{-f_3-\lam |v|^2} \bar{\p} \left( \lam \bar{v}\right) \right)  +  e^{-f_3 - \lam|v|^2}\left( \mf{s} \wedge \bar{\mf{v}} + e^{-f_3} \mf{s}  \w \bar{\mf{s}} \frac{s}{v}\left( e^{-2\lam |v|^2} -1 \right) \right) = 0\, ,
\label{Dmerooffdiag}
\ee
while the vanishing of the diagonal components amounts to
\be
\label{Dmerodiag}
iJ \wedge \p\bar\p f_3  +   e^{-f_3 - 2\lam|v|^2} \left(\frac{1}{2} \mf{s} \w \bar{\mf{s}} + |s|^2 i J \w \p (\lam v) \w \bar{\p} (\lam\bar{v}) \right) + e^{-2f_3} \frac{1}{2} \mf{s} \w \bar{\mf{s}} \left| \frac{s}{v} \right|^2  \left( e^{-2\lam |v|^2} -1 \right)^2 = 0 \, .
\ee
Although they look quite formidable, one can simplify these equations in certain limits. For instance, if we consider eq.\eqref{Dmerooffdiag} for small values of $s$ we can neglect the cubic term in the lhs and recover
\be
 e^{-f_3 - \lam|v|^2} \left[2iJ \wedge \left(\p\bar{\p}(\lam s \bar{v})- s \p (f_3 + \lam |v|^2)\wedge \bar{\p} (\lam \bar{v})\right) + \mf{s} \wedge \bar{\mf{v}}  \right] = 0\, .
\label{linDmerooffdiag}
\ee
This is nothing but the linearised D-term equation $J \wedge \p_{\langle A \rangle} a - \frac{1}{2} [\langle {\Phi} \rangle^\dag, \varphi] = 0$ imposed in the literature to solve for the internal wavefunction profile of fields at matter curves, with $\langle \Phi\rangle$, $\varphi$ the pieces of \eqref{uPhim} at zeroth and linear order in $\mf{s}$, respectively, and similarly for $\langle A \rangle$, $a$ in \eqref{uAm}. Notice that the prefactor in \eqref{linDmerooffdiag} essentially localises the equation along $\Sig = \{v=0\}$, so we may focus on a tubular neighbourhood around the self-intersection locus, as done in local wavefunction computations. Note as well that \eqref{Dmerooffdiag} is a complex equation, so together with \eqref{Dmerodiag} we have three real equations for the two real unknowns $f_3$ and $\lam$. One may see this as a limitation of our Ansatz \eqref{Bxi} and \eqref{xis}, that could be generalised to solve for the most general set of equations. Nevertheless, one may still find solutions with this Ansatz if near $\Sig$ one imposes
\be
J \w  \p \left( s e^{-f_3} \lam \bar{\p} \bar{v}\right)   = 0\, , 
\label{conds}
\ee
after which \eqref{Dmerooffdiag} becomes a real equation. In fact, under these assumptions the dependence of $s$ disappears from \eqref{linDmerooffdiag}, and one obtains a much simpler equation. In particular one may connect with the ultra-local wavefunction results by considering a neighbourhood around a point $p \in \Sig$, and approximating the metric on $S$ to be flat and the 7-brane worldvolume flux to be constant on that neighbourhood. More precisely, if locally we have $v = m_x x$, $J = \frac{i}{2} (dx \wedge d\bar{x} + dy \wedge d\bar{y})$ and $F_3 = \frac{i}{2} (M_x dx \wedge d\bar{x} + M_y dy \wedge d\bar{y})$, \eqref{linDmerooffdiag} is solved by a constant $\lam$ of the form
\be
|m_x|^2 \lam = - \frac{M_x}{2} \pm \sqrt{\left(  \frac{M_x}{2}\right)^2 + |m_x|^2}\, ,
\ee
which reproduces the corresponding local wavefunction solutions, cf. eqs.(2.27) and (2.29) of \cite{Aparicio:2011jx}. Notice that in this particular case a constant $\lam$ implies, through the first condition in \eqref{conds}, that $s$ only depends on the coordinate $y$ of $\Sig$. In this sense, this simple local setup reproduces one of the assumptions of the defect schemes section \ref{s:defects}. In the following we will see how to make this connection more precise and how, by taking the appropriate limit, one can connect the Hitchin D-term equation \eqref{Dmerodiag} with the defect D-term equation \eqref{DSupm}.

\subsection{The defect limits}

As it is clear from the unitary gauge profile for $\Phi$ and $\mb{A}$, the $SU(3)$ Hitchin system above localises the charged fields $s_1^c$ and $s_2$ along the self-intersection curve $\Sig$. In a limit in which such localisation can be approximated by a delta function, one would expect that a defect system should be recovered, and the BPS equations of the $SU(3)$ Hitchin system should become the BPS equations of the meromorphic scheme. In general, one would expect that such a limit is obtained when the intersection slope of the two 7-branes becomes infinite. As we will now see, there are in fact two ways to attain such a limit and recover the defect system. One of them corresponds to increasing the vev of the holomorphic deformation field $v$, and the other to decreasing the overall volume of the four-cycle $S$.

\subsubsection*{The small volume limit}

Let us assume that we have found a solution for the above $SU(3)$ Hitchin system and consider its behaviour under the following rescaling of the four-cycle metric:
\be
|g_S| \raw a^2 |g_S|
\label{limita}
\ee
with $a \in \IR$. As we perform this rescaling the wavefunction profiles for $\Phi$ and $\mb{A}$ are modified, since
\be
\lam |v|^2 = \frac{\lam_\star |v|^2}{\sqrt{|g_S|}}\, \longrightarrow \, \frac{1}{a} \frac{\lam_\star |v|^2}{\sqrt{|g_S|}} = \frac{1}{a} \lam |v|^2\, .
\label{limexp}
\ee
Taking the limit $a \raw 0$ one for instance finds \cite{Cecotti:2010bp}
\be
e^{-\lam |v|^2} \  \stackrel{a\raw 0}{\longrightarrow} \ 2\left[1 - H(|v|^2) \right] \equiv 1 - H_\Sig \, ,
\label{limaxig}
\ee
where $H$ is the Heaviside step function, using the half-maximum convention in which $H(0)=\oh$ and equal to $1$ everywhere else, and we have assumed that $\lam \neq 0$ everywhere in $S$. By \eqref{limaxig}, $H_\Sig$ is a function that vanishes on $\Sig$ and is equal to $1$ everywhere else on $S$. As a consequence
\bea\nonumber
\Phi_{a\raw0} & = &  
\frac{1}{3}\left(
\begin{array}{ccc}
\mf{v} & 0 & 0 \\
0  & \mf{v} &0 \\
0 & 0 & - 2\mf{v}
\end{array}
\right) 
+
\left(
\begin{array}{ccc}
0 & 0 & 0 \\
\frac{s_1^cs_2}{v^2}\mf{v}   & 0 &0 \\
0 & 0 & 0
\end{array}
\right) e^{-f_3} H_\Sig \\ 
& + &
\left(
\begin{array}{ccc}
 0 & 0 & 0 \\
 0 & 0 & e^{-\frac{f_3}{2} + \frac{f_8}{2}} \mf{s}_2  \\
e^{-\frac{f_3}{2} - \frac{f_8}{2}} \mf{s}_1^c  & 0 & 0
\end{array}
\right) \left[1 - H_\Sig \right]
\, .
\label{Philima}
\eea
Notice that only the first line of \eqref{Philima} survives away from  $\Sig$, while the second line is fully localised on top of $\Sig$ as the corresponding, defect fields in the meromorphic scheme. The surviving off-diagonal component is very suggestive in the sense that, again away from $\Sig$, it corresponds to the naive solution to the meromorphic defect equation \eqref{F1m}. 

Now, considering the gauge field in this limit, we have that
\be
 e^{-\lam |v|^2} \bar{\p} \left( \lam \bar{v}\right) \ \longrightarrow  \ e^{-\frac{\lam}{a} |v|^2} \bar{\p} \left( \frac{\lam}{a} \bar{v}\right) \ \stackrel{a\raw 0}{\longrightarrow} \ \bar{\p} \bar{v} \, \pi \de^{(2)} (v)\, ,
\ee
with $\de^{(2)} (v)$ the two-dimensional Dirac delta function with support on $\Sig$. One then finds
\be
i \mb{A}^{(0,1)}_{a\raw0} \, =\, 
\oh H_3  \bar\p f_3 + \oh H_8 \bar\p  f_8 + 
\left(
\begin{array}{ccc}
 0 & 0 & 0 \\
 0 & 0 &  - e^{-\frac{f_3}{2} + \frac{f_8}{2}} s_2 \\
e^{-\frac{f_3}{2} - \frac{f_8}{2}} s_1^c & 0 & 0
\end{array}
\right)  \bar{\p} \bar{v}\, {\pi} \de^{(2)} (v)\,  \, ,
\label{Alima}
\ee
again finding that the profile for the fields $s_1^c, s_2$ is localised on top of $\Sig$, now in the form of a $\delta$-function. Putting these two results together and using the identities
\be
\frac{2}{v} \bar{\p} H(|v|^2) =  \bar{\p}\left(\frac{1}{v}\right) = \pi \de^{(2)} (v) \bar{\p} \bar{v} \, ,
\label{Heavi}
\ee
one can see that the F-terms vanish identically. This is to be expected, since the field-space direction that we are taking to reach this limit does not affect the F-term equations of the Hitchin system. The correct way to extract the F-term \eqref{F1m} is to look at the $\mg{su}(3)$ Hitchin system from the viewpoint of the $\mg{su}(2)$ subalgebra of the corresponding defect scheme. Indeed, one may always rewrite \eqref{uPhim} and \eqref{uAm} as
\be
\Phi \, =\,  \frac{1}{3} \mf{v} \Id_3 +
\left(
\begin{array}{cc}
\Phi_{\mg{su}(2)} & 0\\
0 & - \mf{v}
\end{array}
\right)
+
\Phi_{\rm def}
\label{Phi32}
\ee
and
\be
i \mb{A}^{(0,1)} \, =\, 
 \left(
\begin{array}{cc}
 i  \mb{A}_{\mg{su}(2)} & 0  \\
 0 & - \frac{1}{3} \bar\p f_8
\end{array}
\right)
 +
 i\mb{A}_{\rm def}\, ,
\label{A32}
\ee
where, after the rescaling \eqref{limita},
\be
\Phi_{\mg{su}(2)} = 
e^{-f_3} \left(
\begin{array}{cc}
0 & 0  \\
1 & 0
\end{array}
\right)  \frac{s_1^c s_2}{v^2} \mf{v}  \left(1- e^{-2\frac{\lam}{a} |v|^2 } \right)\, , \qquad 
i \mb{A}^{(0,1)}_{\mg{su}(2)} = 
 \oh \bar\p f_3 T_3 + \frac{1}{6} \bar\p f_8 \mf{1}_2 \, ,
\label{su2s}
\ee
and
\bea
\Phi_{\rm def} & = &
\left(
\begin{array}{ccc}
 0 & 0 & 0 \\
 0 & 0 & e^{-\frac{f_3}{2} + \frac{f_8}{2}} \mf{s}_2  \\
e^{-\frac{f_3}{2} - \frac{f_8}{2}} \mf{s}_1^c  & 0 & 0
\end{array}
\right)   e^{-\frac{\lam}{a} |v|^2}\, ,\\
i\mb{A}_{\rm def} &= &
\left(
\begin{array}{ccc}
 0 & 0 & 0 \\
 0 & 0 &  - e^{-\frac{f_3}{2} + \frac{f_8}{2}} s_2 \\
e^{-\frac{f_3}{2} - \frac{f_8}{2}} s_1^c & 0 & 0
\end{array}
\right)    e^{-\frac{\lam}{a} |v|^2} \bar{\p} \left( \frac{\lam}{a} \bar{v}\right) \, .
\eea
In terms of these quantities, the F-term \eqref{Fterm1} of the $\mg{su}(3)$ Hitchin system reads
\be
\bar{\p}_\mb{A} \Phi = 
\left(
\begin{array}{cc}
\bar{\p}_{\mb{A}_{\mg{su}(2)}} \Phi_{\mg{su}(2)} & 0\\
0 & 0
\end{array}
\right)
-
\Phi_{\rm def} \wedge \bar{\p} \left( \frac{\lam}{a} \bar{v}\right) v
-i \left[\mb{A}_{\rm def},
{\small \left(
\begin{array}{ccc}
 0 & &  \\
 & 0 & \\ 
 & & -\mf{v}
\end{array}
\right)}\right]
-i [\mb{A}_{\rm def}, \Phi_{\rm def}]\, .
\ee
One can check that
\be
\Phi_{\rm def} \wedge \bar{\p} \left( \frac{\lam}{a} \bar{v}\right) v
+ i \left[\mb{A}_{\rm def},
{\small \left(
\begin{array}{ccc}
 0 & &  \\
 & 0 & \\ 
 & & -\mf{v}
\end{array}
\right)}\right] \equiv 0\, ,
\ee
and that 
\be
i [\mb{A}_{\rm def}, \Phi_{\rm def}] = 
\left(
\begin{array}{ccc}
 0 & 0& 0 \\
\Xi & 0 & 0\\ 
 0 & 0 & 0
\end{array}
\right)\, , \qquad  \Xi = - 2 s_1^c \mf{s}_2 \wedge  \bar{\p} \left( \frac{\lam}{a} \bar{v}\right)\, e^{-f_3-2\frac{\lam}{a} |v|^2}\, .
\label{defXi}
\ee
Therefore, satisfying the F-terms for the $\mg{su}(3)$ Hitchin system amounts to imposing that
\be
\bar{\p}_{\mb{A}_{\mg{su}(2)}} \Phi_{\mg{su}(2)}  = \Xi\, ,
\label{F1su2Xi}
\ee
and taking the limit $a \raw 0$ one obtains
\be
\Xi\; \stackrel{a\raw0}{\longrightarrow}\; \Xi_{0}= - \pi e^{-f_3} s_1^c {\bf s_2}   \wedge \bar{\p} \bar{v} \, \de^{(2)}(v) \, .
\label{Xilima}
\ee
The defect F-term \eqref{F1m} can be recovered from \eqref{F1su2Xi} and \eqref{Xilima} as follows. Recall that, due to the presence of the delta function in eq. \eqref{Xilima}, we can split the $(2,0)$-form ${\bf s_2}$ into a $(1,0)$-form longitudinal to the curve $\Sig$ and a $(1,0)$-form transverse to it. Close to $\Sig$, the latter has the canonical form $\p v$, hence we can write \eqref{Xilima} as
\be\label{Xilima2}
\Xi_{0} = 2\pi i e^{-f_3}s_1^c s_2\wedge \delta_\Sig\,,
\ee
where $s_2\in H^{0}(S,\cL_3^{-1}\times K_S)$, $s_1^c\in H^1(S,\cL_3^{-1})\simeq H^1(S,\cL_3\times K_S)^*$ by Serre duality on $S$, and we have used that
\be\label{delta_Sigma}
\delta_\Sig = \frac{i}{2}\delta^{(2)}(v)\p v\wedge \bar{\p}\bar{v}\,.
\ee
Since only the profile of $s_1^c s_2$ on top of $\Sig$ matters in \eqref{Xilima2}, we can write $s_2\in H^{0}(\Sig,\hat{\cL}_3^{-1}\times K_S|_\Sigma)\simeq H^0(\Sig,\hat{\cL}_3^{-1}\times K_\Sigma^{1/2})$ due to $K_S|_\Sig\simeq K_\Sigma^{1/2}$, and $s_1^c\in H^1(\Sigma,\hat{\cL}_3\times K_S|_\Sig)^*\simeq H^0(\Sig, \hat{\cL}_3^{-1}\times K^{1/2}_\Sig)$ by Serre duality on $\Sig$. Therefore, we have shown that $s_1^cs_2\in H^0(\Sig,\hat{\cL}_3^{-2}\times K_\Sigma)$ and thus it transforms correctly as a $(1,0)$-form on $\Sigma$ with the same charge of the field $p$ along the $T_-$ generator. To conclude the proof that \eqref{F1su2Xi} is indeed equivalent to \eqref{F1m}, we need to go back to the holomorphic gauge, but now from the $\mg{su}(2)$ perspective, whereby, remarkably, we just have the standard split connection \eqref{su2s}. This operation just removes the $f_3$-dependent factors, and in the $a\to0$ limit we get exactly \eqref{F1m} upon the identifications
\be
\sig_1^c =  \sqrt{2\pi i}\, s_1^c \,,\qquad \sig_2 =  \sqrt{2\pi i}\, s_2\,.
\ee
There is a small caveat here: In the limit $a\to0$ we are able to reach an holomorphic gauge everywhere on $S$, except on the curve $\{v=0\}$. This is because the limiting $\mg{su}(2)$ Higgs field, in the holomorphic gauge, looks like
\be
\Phi_{\mg{su}(2)}^{\rm h}|_{a\to 0} =  \frac{s_1^c s_2}{v^2} \mf{v}  H_\Sig T_-\,,
\ee
which gets a dependence on $\bar{v}$ from the step function $H_\Sig$. This is not a problem, because, as we are going to explain below, the defect picture is an appropriate description of the system outside of a tubular neighbourhood of $\Sig$. In the limit, such a neighbourhood shrinks and eventually coincides with $\Sig$, making the defect picture reliable everywhere except on $\Sig$.

\subsubsection*{The large angle limit}

Even if the small volume limit reproduces the F-terms of the meromorphic scheme, the D-term equation \eqref{Dterm} cannot be trusted in the regime where it applies. One may nevertheless conceive a second limit, which amounts to increasing the vev of the intersection field $v$
\be
v \, \raw \, b\, v
\label{limitb}
\ee
with $b \in \IR$, while keeping the four-cycle metric fixed. Taking $b \raw \infty$ will ultra-localise the fields at the self-intersection $\Sig$, and so one would expect to recover again the delta function behaviour of the defect scheme. This time, because we are at large volume, it makes sense to try and solve the D-terms as we vary $b$. It is particularly important that the off-diagonal D-terms in \eqref{Dtermmero} are identically satisfied as we move along \eqref{limitb}, because these correspond to the D-term potential for massive fields at the self-intersection. Since such fields are assumed to be very massive and completely integrated out in the regime where the defect picture is valid, one would never attain the defect limit unless one sets their D-terms to zero. For doing so, let us take the simplifying assumption $|s| = |s_1^c| = |s_2|$ that takes us to  \eqref{Dtermmero} and assume that we have a configuration such that can find a solution of both D-term equations $C= D = 0$ with our Ansatz \eqref{xis}. Performing the rescaling \eqref{limitb}, eq.\eqref{Dmerooffdiag} transforms as
\be\nonumber
2iJ \wedge \left(b\p\bar{\p}(\lam s \bar{v})- s \p (f_3 + b^2 \lam v\bar{v})\wedge b \bar{\p} (\lam \bar{v})\right) + b \mf{s} \wedge \bar{\mf{v}} + e^{-f_3} \mf{s}  \w \bar{\mf{s}} \frac{s}{bv}\left( e^{-2\lam b^2|v|^2} -1 \right)  = 0\, ,
\label{Dmoffdiaglimitb}
\ee
where we have discarded overall exponential factors. In the limit $b \raw \infty$, we will  be able to find a solution only if $\lam$ also scales with $b$ in the following form
\be
\lam \, \raw b^{-1} \lam
\label{limbl}
\ee
where this should be interpreted as a rescaling of the function $\lam_\star$ in \eqref{deflam} and not of the metric factor therein. Notice that the rescalings \eqref{limitb} and \eqref{limbl} have the same combined effect on $\lam|v|^2$ as in \eqref{limexp}, with the replacement $a^{-1} \raw b$, so as in the previous limit we expect a strong localisation for the intersection fields as we reach $b \raw \infty$. This time, however, we also need to consider the behaviour of non-holomorphic data like kinetic terms. Indeed, the kinetic-term integrand for the intersection fields scale like 
\bea\nonumber
i J \wedge [\mb{A}_{\rm def}, \mb{A}_{\rm def}^\dag] + \oh [\Phi_{\rm def}, \Phi_{\rm def}^\dag] &  = & - \left(i |s|^2  J \wedge \p (\lam v) \wedge \bar{\p} (\lam \bar{v}) + \oh \mf{s} \wedge \bar{\mf{s}} \right) e^{-f_3 - 2\lam |v|^2}\\ \nonumber
& \raw & - \left(i |s|^2  J \wedge \p (\lam v) \wedge \bar{\p} (\lam \bar{v}) + \oh \mf{s} \wedge \bar{\mf{s}} \right) e^{-f_3 - 2b\lam |v|^2}\ ,
\eea
and so the kinetic terms will vanish in the limit $b \raw \infty$. This can be fixed by rescaling the normalisation factor of the fields at the intersection, which in practice amounts to 
\be
s \, \raw \, b^{1/2} s \, .
\label{limbs}
\ee
Compared to \eqref{limita}, the effect of the combined rescaling \eqref{limitb}, \eqref{limbl} and \eqref{limbs} on $\Phi$ and $\mb{A}$ is slightly different. Nevertheless, the effect on \eqref{defXi} is similar, and so we recover the same limiting behaviour \eqref{Xilima} that reproduces the F-terms of the meromorphic scheme. 

Let us now consider the D-term equation, and in particular the non-Cartan components of \eqref{Dtermmero}. After taking the limit $b \raw \infty$ most of its terms vanish automatically, except one proportional to
\be
\left. J \w  \p \left( s e^{-f_3} \lam \bar{\p} \bar{v}\right) \right|_\Sig \, .
\label{condsig}
\ee
%
%
As pointed out before, the vanishing of this quantity is what allows to convert $D=0$ into a real equation and to find solutions for the D-term equations within the Ansatz \eqref{xis}. As we are using such an Ansatz to connect with the defect scheme it seems reasonable that, by consistency, we should restrict to configurations where \eqref{condsig} vanishes. 

Finally, the diagonal component of \eqref{Dtermmero} scales as
\be
\label{Dmerodiagb}
iJ \wedge \p\bar\p f_3  +   e^{-f_3 - 2b \lam|v|^2} b |s|^2 \left(\frac{J^2}{4 \sqrt{|g_S|}}  +  i J \w \p (\lam v) \w \bar{\p} (\lam\bar{v}) \right) + e^{-2f_3} \frac{1}{2} \mf{p} \w \bar{\mf{p}}   \left( e^{-2b \lam |v|^2} -1 \right)^2 ,
\ee
where we have defined $\mf{p} =  \frac{s^2}{v^2} \mf{v}$. Taking the limit $b \raw \infty$, and assuming that in a neighbourhood of $\Sig$ the following relation holds
\be
 2i \lam^2 \sqrt{|g_S|} J \wedge \p v \wedge \bar{\p} \bar{v}  = \oh J^2 \, ,
 \label{relJ}
\ee
we recover the following D-term equation
\be
\label{Dmerodiaglimb}
- iJ \wedge \p\bar\p f_3    =  e^{-2f_3} \frac{1}{2} \mf{p} \w \bar{\mf{p}}\, H_\Sig +  \lam_\star  \frac{e^{-f_3}}{ \sqrt{|g_S|} } |s|^2 \, 2 \pi \de_\Sig \wedge J \, ,
\ee
where we have used the relation \eqref{delta_Sigma}

We then recover the meromorphic-scheme D-term equations  \eqref{DSupm} with $\mf{m} \equiv 0$, upon identifying $h_\CL =  h_3 = e^{f_3}$, $ \sqrt{|g_S|} = h_\Sig^{1/2}$ and $4\pi \lam_\star |s|^2 = |\sig|^2$. In fact, strictly speaking we only reproduce the defect equations away from the self-intersection locus $\Sigma$, due to the appearance of $H_\Sig$ in \eqref{Dmerodiaglimb}. This is nevertheless consistent with the regimes in which  the $\mg{su}(3)$ Hitchin system and the $\mg{su}(2)$ system with defects are reliable descriptions.

Indeed, the regular $\mg{su}(3)$-Hitchin-system description that we are using should only be valid in regions of $S$ where $|v|$ is small compared to the string scale, and beyond that the Hitchin description should only be strictly valid for the $\mg{su}(2)$ sector. The degrees of freedom that are left out from the Hitchin system are those outside of $\mg{su}(2)$, and in particular the non-Cartan entries of $\Phi$ and $\mb{A}$ that include the fields localised at the self-intersection curve $\Sig = \{v=0\}$ and their massive replicas. As we increase the vev of $v$ through the rescaling \eqref{limitb}, this region of validity narrows down as a tubular region around $\Sig$. This limits the computation of certain non-holomorphic 4d couplings by dimensional reduction, namely those whose integrand does not converge sufficiently fast in that region. This does not seem to be a problem for the kinetic terms of the light localised modes $s_1^c$, $s_2$ if we perform the rescaling \eqref{limbs}, but it should affect the kinetic terms of massive modes in the same sector that have a mass comparable to the string scale. In order to  correctly integrate out these massive modes one needs to solve their corresponding D-term equations, which are encoded in the non-Cartan D-term equation \eqref{Dmerooffdiag}. Remarkably, solving these equations at an intermediate stage of the large angle limit implies imposing the relations \eqref{conds} and \eqref{relJ} in the corresponding tubular neighbourhood. 

This region of validity is somewhat opposite for the defect description. For instance, let us look at the entry of $\Phi$ that gives rise to the   F-term pole, namely at the piece
\be
  \frac{s}{v} \mf{s} \left(1- e^{-2b\lam |v|^2} \right)\, .
 \label{volcano}
\ee
Whenever $|g_S|^{-1/2}|v|^2 \gg (\lam_\star b)^{-1}$  this piece reduces to the meromorphic (2,0)-form $\frac{s}{v} \mf{s}$, so at this distance from $\Sig$ it looks like the $\mg{su}(2)$ 7-brane sector develops a pole. In fact, as discussed above, at this distance the Hitchin system is only good to describe the $\mg{su}(2)$ subsector of $\mg{su}(3)$. Therefore, it is more useful to think of the non-Cartan fields $s$ as a separate sector, as the defect picture does. As we enter the region $|g_S|^{-1/2}|v|^2 \leq (\lam_\star b)^{-1}$  the Hitchin system description starts being reliable to describe the $\mg{su}(3)$ system. Then we see that the pole-like behaviour $\frac{s^2}{v}$ starts being softened by the exponential, and that the (2,0)-form \eqref{volcano} actually vanishes at $v=0$. The norm of \eqref{volcano} looks like a volcano-shaped profile: From far away it seems to develop a pole at $v=0$, but close to $\Sig$ there is a turning point that makes the function go down to zero. In the limit $b\raw \infty$ this becomes the function $\left| \frac{s^2}{v}\right|^2 H_\Sig$ that appears in \eqref{Dmerodiaglimb}. The $\mg{su}(2)$ modes whose profile is mostly outside of this region will see a pole, because their coupling is given by an integral that does not care much about the interior of the volcano. It is for those modes that the defect picture is useful. In the strict limit $b \raw \infty$ this set amounts to essentially all $\mg{su}(2)$ modes, in agreement with the fact that $H_\Sig$ is {different from $1$ on a space of zero measure} and its presence does not affect the integrals that give rise to the 4d D-term potential.

\subsection{The holomorphic scheme}

Let us now consider the regular $SU(3)$ Hitchin system that is related to the holomorphic scheme in the self-intersecting curve $S$. As many of the ingredients are similar to the meromorphic scheme, our discussion will be more sketchy for this case. We start from the following holomorphic Higgs field
\be
\Phi^{\rm h} \, =\,  \frac{1}{3} 
\left(
\begin{array}{ccc}
{\bf v} & 0 & 0 \\
0 & {\bf v} & 0 \\
0 & 0 & -2{\bf v}
\end{array}
\right)
+
\left(
\begin{array}{ccc}
0 & \mf{m} & 0 \\
 0 & 0 & {\bf s}_2 \\
0 & 0 & 0
\end{array}
\right)\, ,
\label{hPhih}
\ee
with $\mf{s}_2 \in H^{2,0} \left({\CL}_3^{-1} \otimes {\CL}_8\right)$ and $\mf{m} \in H^{2,0} \left({\CL}_3^{2}\right)$. We choose a complexified gauge transformation of the form
\be
B \, =\,
\left(
\begin{array}{ccc}
e^{f_3/2+f_8/6} & 0 & 0 \\
0  & e^{-f_3/2+f_8/6} & 0  \\
0 & 0 & e^{-f_8/3}
\end{array}
\right)
\cdot
\left(
\begin{array}{ccc}
1 & 0 & -\xi_2 \xi_m \\
0  & 1 & - \xi_2  \\
0 & 0 & 1
\end{array}
\right)
\label{Bxih}
\ee
where $\xi_2$ is given by \eqref{xis} and 
\be
\xi_m \, =\, \frac{m}{v}  \, \left(e^{-\mu |v|^2 }-1 \right)\, ,
\label{xim}
\ee
with $\mu = |g_S|^{-1/2} \mu_\star$ and $\mu_\star$ another global real function on $S$. The Higgs field in the unitary frame is now given by
\be
\Phi \, =\,  \frac{1}{3}\mf{v} \Id_3 +
\left(
\begin{array}{ccc}
0 & e^{f_3} \mf{m} & - e^{\frac{f_3}{2} + \frac{f_8}{2}} e^{-\mu |v|^2}  \left(e^{-\lam |v|^2 }-1 \right) \frac{s_2}{v} \mf{m}\\
0 & 0 &  e^{-\frac{f_3}{2} + \frac{f_8}{2}}  e^{- \lam  |v|^2} \mf{s}_2 \\
0 & 0 & -\mf{v}
\end{array}
\right)\, ,
\label{uPhih}
\ee
while the gauge connection is given by
\bea\label{uAh}
i \mb{A}^{(0,1)}  & = & H_1  \bar\p f_3 + \frac{1}{\sqrt{3}} H_2 \bar\p  f_8 - e^{-\frac{f_3}{2} + \frac{f_8}{2}} e^{-\lam |v|^2} \bar{\p} \left( \lam \bar{v}\right) s_2 \, \eps_2  \\ \nonumber
& - & e^{\frac{f_3}{2} + \frac{f_8}{2}} \left[ e^{-\lam |v|^2}  \left(e^{-\mu |v|^2 }-1 \right)\bar{\p} \left( \lam \bar{v} \right) + e^{-\mu |v|^2}  \left(e^{-\lam |v|^2 }-1 \right)\bar{\p} \left( \mu \bar{v} \right)\right] \frac{m s_2}{v} \, \eps_{12} \, ,
\eea
where we have used the notation of Appendix \ref{ap:su3} for the algebra generators $\{H_1, H_2, \eps_2, \eps_{12}\}$. These two expressions simplify considerably in the small volume limit:
\be
\Phi_{a\raw0}\, =\,  \frac{1}{3}\mf{v} \Id_3 +
\left(
\begin{array}{ccc}
0 & e^{f_3} \mf{m} & 0 \\
0 & 0 &  e^{-\frac{f_3}{2} + \frac{f_8}{2}} \left[ 1 - H_\Sig\right] \mf{s}_2 \\
0 & 0 & -\mf{v}
\end{array}
\right)\, ,
\label{Philimah}
\ee
\be
i \mb{A}^{(0,1)}_{a\raw0} \, =\, 
\oh H_3  \bar\p f_3 + \oh H_8 \bar\p  f_8 + 
\left(
\begin{array}{ccc}
 0 & 0 & 0 \\
 0 & 0 &  - e^{-\frac{f_3}{2} + \frac{f_8}{2}} s_2 \\
0 & 0 & 0
\end{array}
\right)  \bar{\p} \bar{v}\, {\pi} \de^{(2)} (v)\,  \, ,
\label{Alimah}
\ee
again displaying a split {connection} for the $\mg{su}(2)$ subalgebra and field localisation for $s_2$. The main difference with respect to the meromorphic case is that now the dependence on the defect field $s_2$ is completely localised on $\Sig$, and as a consequence no pole arises. Indeed, performing the split of eqs.\eqref{Phi32} and \eqref{A32} and repeating the computation below them, one again finds the result \eqref{F1su2Xi}, but now with $\Xi =0$ due to the absence of a vev for $s_1^c$.

Let us now analyse the D-terms, whose structure in this case is general
\be
J \wedge \mathbb{F} +\frac{1}{2} [\Phi, \Phi^\dagger] = 
\left(
\begin{array}{ccc}
 C_1 \quad & F &  {E} \\
 \bar{F} \quad  & C_2 &  D \\
  \bar{E} \quad  &  \bar{D} & - C_1 - C_2
\end{array}
\right)\, .
\label{Dtermholo}
\ee
The D-term equation then amounts to three complex and two real equations, while our Ansatz contains four unknown functions: $\{f_3, f_8, \lam, \mu\}$. To solve the D-term equations within this Ansatz one then needs to make further assumptions. For instance, let us consider the condition $D=0$, which reads
\bea\label{DholoD}
& & 2 i J \w  \p \left( s_2  e^{-f_3+f_8 -\lam |v|^2} \bar{\p} \left( \lam \bar{v}\right) \right)  + \\ \nonumber & & +\,   e^{-f_3+f_8} \left( e^{ - \lam|v|^2} \mf{s}_2 \wedge \bar{\mf{v}} + e^{-2f_3-\mu|v|^2} \mf{m}  \w \bar{\mf{m}} \frac{s_2}{v}\left( e^{-2\lam |v|^2} -1 \right) \right) = 0\, .
\eea
If again we impose \eqref{conds} (now with the replacement $f_3 \raw f_3 - f_8$) in a neighbourhood of $\Sig$, this complex equation becomes a real one. In fact, upon performing the rescaling 
\be
v \, \raw \, bv\, ,  \qquad \lam\, \raw \, b^{-1} \lam\, , \qquad \mu\, \raw \, b^{-1} \mu\, , \qquad s_2 \, \raw \, b^{1/2} s_2\, , \qquad m\, \raw \, m\, , 
\label{limitbh}
\ee
and taking the large angle limit $b\raw \infty$, satisfying \eqref{DholoD} amounts to imposing \eqref{conds} on top of $\Sig$, in analogy with the corresponding non-Cartan equation in the meromorphic scheme. Regarding the condition $E=0$, which is equivalent to
\bea\nonumber
& & 2 i J \w  \p  \left[ \frac{ms_2}{v}  e^{f_3+f_8} \left( e^{-\lam |v|^2}  \left(e^{-\mu |v|^2 }-1 \right)\bar{\p} \left( \lam \bar{v} \right) + e^{-\mu |v|^2}  \left(e^{-\lam |v|^2 }-1 \right)\bar{\p} \left( \mu \bar{v} \right)\right) \right]  \\  & & + \, e^{f_3+f_8}  e^{-\mu |v|^2}  \left(e^{-\lam |v|^2 }-1 \right) \frac{s_2}{v} \mf{m} \wedge \bar{\mf{v}}  = 0\, ,
\label{DholoE}
\eea
one can see that all the terms vanish as we take the  large angle limit. Something similar happens for the condition $F=0$:
\bea\nonumber
& & 2 i J \wedge  |{s}_2|^2 \frac{m}{v}  e^{f_8-\lam |v|^2} {\p} \left( \lam {v}\right)\wedge  \left[ e^{-\lam |v|^2}  \left(e^{-\mu |v|^2 }-1 \right)\bar{\p} \left( \lam \bar{v} \right) + e^{-\mu |v|^2}  \left(e^{-\lam |v|^2 }-1 \right)\bar{\p} \left( \mu \bar{v} \right)\right] \\
& &  =\,  e^{f_8-(\mu + \lam) |v|^2}  \left(e^{-\lam |v|^2 }-1 \right) \frac{s_2}{v} \mf{m} \wedge \bar{\mf{s}}_2 \, ,
\label{DholoF}
\eea
Indeed, one can check that both sides of the equation vanish as we take the limit  $b \raw \infty$. Finally, we have two D-term equations corresponding to the Cartan generators of $\mg{su}(3)$. The condition $C_1 = 0$ amounts to imposing
\bea\label{DholoC1}
& &2 i J \wedge \p\bar{\p} \left(\frac{1}{3} f_8  + f_3 \right) + 2 i e^{f_3 + f_8} \left| m s_2\right|^2 J \wedge \zeta \wedge \bar{\zeta} \\ \nonumber
& & = e^{2f_3} \mf{m} \wedge \bar{\mf{m}} + e^{f_3 + f_8} \left(e^{-\lam |v|^2 }-1 \right)^2  e^{-\mu |v|^2}  \left| \frac{s_2}{v}\right|^2 \mf{m} \wedge \bar{\mf{m}}
\eea
where
\be
\zeta = \frac{1}{\bar{v}}  \left[ e^{-\lam |v|^2}  \left(e^{-\mu |v|^2 }-1 \right){\p} \left( \lam {v} \right) + e^{-2\mu |v|^2}  \left(e^{-\lam |v|^2 }-1 \right){\p} \left( \mu{v} \right)\right] \, .
\ee
The equation $C_2 =0$ reads in turn
\bea\label{DholoC2}
& &2 i J \wedge \p\bar{\p} \left(\frac{1}{3} f_8 - f_3 \right) - 2 i e^{-f_3 + f_8} e^{-2\lam |v|^2} \left| s_2\right|^2 J \wedge \p \left(\lam v \right)  \wedge \bar{\p} \left( \lam \bar{v} \right) \\ \nonumber
& & = \, e^{-f_3 + f_8} e^{-2\lam |v|^2} \mf{s}_2 \wedge \bar{\mf{s}}_2 - e^{2f_3} \mf{m} \wedge \bar{\mf{m}}\, .
\eea
Upon taking the large angle limit, $\zeta$ vanishes, as does the second term in the rhs of \eqref{DholoC1}. In addition, \eqref{DholoC2} simplifies after using the relation \eqref{relJ} on top of $\Sig$. We are finally left with 
\bea
- i J \wedge \p\bar{\p} \left(\frac{1}{3} f_8  + f_3 \right) & = & -\oh e^{2f_3} \mf{m} \wedge \bar{\mf{m}}\, ,\\
- i J \wedge \p\bar{\p} \left(\frac{1}{3} f_8  - f_3 \right) & = & \oh e^{2f_3} \mf{m} \wedge \bar{\mf{m}} -  \lam_\star \frac{e^{-f_3 + f_8}}{\sqrt{|g_S|}} |s_2|^2 2 \pi \de_\Sig \wedge J \ ,
\eea
and so we recover \eqref{DSup} and \eqref{DSdown} upon the identifications  $h_\CL =  h_3 = e^{f_3}$, $ \sqrt{|g_S|} = h_\Sig^{1/2}$ and $4\pi \lam_\star |s_2|^2 = |\sig_2|^2$.


\section{Conclusions}
\label{s:conclu}

In this paper we have extended our previous work \cite{Marchesano:2017kke} to analyse global aspects of T-branes in compact K\"ahler surfaces in the presence of defects. Defect fields are essentially always there, in the sense that, in a consistent compactification, a 7-brane stack wrapping a non-trivially embedded surface $S$ will generically intersect at least a second 7-brane stack wrapping $S'$, and such an intersection will typically host some localised fields. The working assumption of \cite{Marchesano:2017kke} was to consider a vanishing vev for such localised fields, so that their presence could be ignored at the level of the 7-brane background on $S$. As a result, the BPS conditions to be satisfied by the background were given by a regular Hitchin system on $S$. In this work we dropped this assumption, and explored which kind of T-brane backgrounds one can construct when the vev of the defect fields is non-vanishing. From the viewpoint of the Hitchin system on $S$, we are now allowing for $\delta$-function sources in its equations. From the point of view of the compactification, we are considering more general bound states of 7-branes than in \cite{Marchesano:2017kke}. 

One of the main motivations for this extension is to see if the obstructions found in \cite{Marchesano:2017kke} to building T-branes in compact surfaces are relaxed in the presence of defects. Particularly interesting are the surfaces of positive Ricci curvature, which are forbidden for the T-brane configurations considered in \cite{Marchesano:2017kke}. This kind of surfaces are specially meaningful in F-theory GUT models, as they can be identified with the four-cycle $S_{\rm GUT}$ hosting all the degrees of freedom charged under the group of grand unification $G_{\rm GUT}$. Even if the T-brane sector of F-theory GUT models lies beyond $G_{\rm GUT}$ and as such it may be located on a different surface, being able to define it over $S_{\rm GUT}$ allows to have a better control over the internal wavefunctions of the different 4d GUT light fields, and therefore on their couplings. 

Here we have found that, when turning on vevs for defect fields, one can indeed construct T-brane backgrounds on surfaces of positive curvature. We have devised two different schemes in which this may happen: One in which the profile for the Higgs field contains poles, thereby involving meromorphic sections of line bundles on $S$, and the other in which all sections are holomorphic. The meromorphic scheme is particularly interesting, in the sense that it is compatible with manifolds of positive curvature and T-brane backgrounds of the form \eqref{NonNilpotentPhiStack}, two of the ingredients used to construct F-theory GUTs with realistic Yukawas \cite{Cecotti:2010bp,Font:2013ida,Marchesano:2015dfa,Carta:2015eoh}. Nevertheless, these backgrounds come with their own constraints, that stem from the Yukawa couplings involving two defect modes and one bulk mode, and that may in some cases forbid the necessary combinations of defect vevs to realise the scenario. In general, pairs of defect modes also have Yukawa couplings with massive bulk modes. In this case the Yukawas do not induce any constraint on the system. Quite on the contrary, they provide a 4d description of why a pole is generated for the Higgs-field profile.

One may compare either the meromorphic or the holomorphic defect schemes with an analogous construction of the kind analysed in \cite{Marchesano:2017kke}, by replacing the 7-brane on the intersecting surface $S'$ with yet another 7-brane on the stack on $S$. One can see that in the presence of similar gauge bundles the effective field theories in both cases are the same, and that the only thing that changes are the necessary conditions for the existence of zero modes. In the case of the defect schemes of section \ref{s:defects} one needs to require the existence of modes localised on $\Sig = S \cap S'$, while in the analogous configuration in \cite{Marchesano:2017kke} one would require the existence of a bulk mode. The latter condition is much more constraining than the former from the viewpoint of the topology of $S$, and that is essentially why the present defect schemes are able to circumvent the no-go result of \cite{Marchesano:2017kke}.

Somehow reversing this strategy, one may take a stack of 7-branes on a surface $S$ with holomorphic deformations, and deform the embedding of one 7-brane to a homotopic surface $S'$. In general $S'$ will intersect the rest of the stack at the self-intersection curve $\Sig$, where fields will be localised. One would expect that, whenever the intersection angle at $\Sig$ is much larger than the typical scale of $S$ one should be able to describe them as defects, and therefore recover a particular case of Hitchin system with defect sources. In section \ref{s:Hitchin} we have made such an intuition precise by describing two limits under which, starting from an ordinary Hitchin system without defects, one can recover the $\de$-function sources of either the above holomorphic or meromorphic defect schemes. The first of the two limits we considered corresponds to decreasing the overall volume of $S$, entering a regime in which the standard D-term equations of the Hitchin system cannot be trusted. Therefore it does not make sense to compare them with the ones of the defect system. It would however be interesting to see if the $\alpha'$-corrected version of these equations \cite{Minasian:2001na,Marchesano:2016cqg} could allow to make contact with the D-term equations with defects also under this limit.

Notice that the setup of section \ref{s:Hitchin} is quite specific, in the sense that it requires that the surface $S$ has holomorphic deformations, and in particular an effective canonical bundle. However, the lessons to be drawn are more general. Indeed, consider two stacks of 7-branes wrapping the surfaces $S$ and $S'$ and intersecting on the curve $\Sig$. Depending on how steep the angle of intersection is, we can define a tubular neighbourhood of $\Sig$ in $S$, in which the symmetry group is enhanced from $G_S$ to $G_\Sig \supset G_S \times G_{S'}$, and the degrees of freedom localised on $\Sig$ are described by a regular Hitchin system with group $G_\Sig$. One should then be able to consider similar limits to those implemented in section \ref{s:Hitchin}, in order to recover a Hitchin system with symmetry group $G_S$ and defect sources on $\Sig$.

The analysis of this paper opens up several interesting avenues for further investigation. As mentioned, Hitchin systems with defects are central in the analysis of \cite{Anderson:2013rka,Anderson:2017zfm}, where the coupling to gravity (and thus the compactification of the internal dimensions) is also considered, though, differently from here, directly at the level of the F-theory lift. While, on the one hand, their investigation is limited to six-dimensional vacua, we only focused, on the other hand, on the compactness of the 7-brane locus, disregarding that of its embedding space. It would therefore be very interesting to try and combine the two approaches, looking at the implications each one has on the other.

In this paper we have restricted our attention to defects involving simple (i.e. order-one) poles. It would be important to extend our analysis to include higher-order poles for the Higgs field, and thus derive global consistency conditions for T-branes gone wild \cite{Anderson:2017zfm}. Also, we only considered defect sources which could be associated to vevs of bifundamental fields: It would be nice to see how our findings generalise to cases where the modes that condense come from tensionless strings \cite{Anderson:2013rka}.

There is an interesting observation related to poles for off-diagonal Higgs-field modes, which may be worth investigating further in the future. While systems with poles for diagonal modes admit an alternative description in terms of recombined 7-branes, configurations featuring poles for off-diagonal modes may be viewed as split sub-loci hosting localised matter which recombine. Let us see how this comes about in our meromorphic scheme, where we have a stack of two D7-branes wrapping the locus $S:\{t=0\}$, a single D7-brane on $S':\{\tau=0\}$, and non-trivial vevs for the modes $\sigma_1^c$, $\sigma_2$ localised at $\Sig=S\cap S'$, inducing a pole for the bulk $p$ mode along the $T_-$ generator of $\mg{sl}(2)$. In the absence of holomorphic $p$ modes, the system can be described by a tachyon condensation process of D9 and anti-D9-branes, with the following tachyon profile:\footnote{See e.g. \cite{Collinucci:2014qfa} where these concepts are reviewed in the context of T-branes.}
\be\label{Tachyon}
T=\left(\begin{array}{ccc} t&0&0 \\ 0&t&\sigma_2 \\ \sigma_1^c&0 &\tau \end{array}\right)\,.
\ee
Looking at how the rank of this matrix jumps, we quickly realise that matter is trapped on the split locus $\{\sigma_1^c\sigma_2=0\}\subset S$. Note how, from this different perspective, the original intersection curve $\Sig$ has lost any significance. If, on the contrary, the topology allows for holomorphic $p$ sections, a non-trivial vev for them would cause a recombination of the above two branches of the matter curve. This is equivalent to switching on a $(2,1)$ entry $p$ in \eqref{Tachyon}, thereby turning the locus hosting matter into the recombined curve $\{p \tau-\sigma_1^c\sigma_2=0\}\subset S$. The same phenomenon takes place if, in addition to the above, an holomorphic $m$ mode (i.e. along the $T_+$ generator) is given a vev, except that, in this case, matter undergoes a further localisation onto the set of points where $m$ vanishes. This alternative picture of the 7-brane system is equivalent as far as holomorphic information is concerned. It would remain to see to what extent this analogy can be carried over to include D-term data.

\bigskip

\centerline{\bf \large Acknowledgments}

\bigskip

RS would like to thank the IFT in Madrid for kind and constant hospitality at various stages of this project. SS would like to thank the Department of Physics of the University of Rome Tor Vergata for hospitality during this project. This work is supported by the Spanish Research Agency (Agencia Estatal de Investigaci\'on) through the grant IFT Centro de Excelencia Severo Ochoa SEV-2016-0597, by the grant FPA2015-65480-P from MINECO/FEDER EU. SS is supported by the FPI grant SVP-2014-068525. RS is supported by the program Rita Levi Montalcini for young researchers (D.M. n. 975, 29/12/2014).

\appendix


\section{Lie algebra conventions}
\label{ap:su3}

In section \ref{ss:compact} we make use of both the generators of the complexified Lie-algebra $\mg{su}(2)_{\mb{C}} = \mg{sl}(2)$ as well as $\mg{su}(3)_{\mb{C}} = \mg{sl}(3)$. Let us therefore summarise here the conventions used for the generators.

\subsection*{$\mg{sl}(2)$ generators}
As already indicated in \ref{ss:compact}, we use the following conventions
\begin{align}
	T_3 = \left(\begin{matrix}
	1 & 0 \\ 0 & -1
	\end{matrix}\right)\,, \qquad T_+ = \left(\begin{matrix}
		0 & 1 \\ 0 & 0
	\end{matrix}\right)\,, \qquad T_- = \left(\begin{matrix}
		0 & 0 \\ 1 & 0
	\end{matrix}\right)\,,
\end{align}
satisfying the commutation relations
\begin{align}
	[T_+, T_-] = T_3\,, \qquad [T_3, T_+] = 2 T_+ \,, \qquad [T_3, T_-] = -2 T_-\,.
\end{align}

\subsection*{$\mg{sl}(3)$ generators}
In the main text our examples were constructed in an $\mg{su}(3)$ background, where we made use both of the generators in the Cartan-Weyl basis as well as in the Chevalley basis, which has only integer structure constants. For convenience we give both bases explicitly here.

We denote the generators in the Cartan-Weyl basis by capital letters. The two Cartan elements are given by
\be
H_1=\frac{1}{2} \left(\begin{array}{ccc} 1 &0&0\\0&-1 &0\\ 0&0&0
\end{array}\right)\,,\qquad H_2=\frac{1}{2\sqrt{3}} \left(\begin{array}{ccc} 1&0&0\\0&1&0\\ 0&0&-2
\end{array}\right)\,,
\label{ap:Cartan38}
\ee
whereas the simple and highest roots are given by
\be
E_1= \frac{1}{\sqrt{2}} \left(\begin{array}{ccc} 0&1&0\\0&0&0\\ 0&0&0
\end{array}\right),\quad E_2= \frac{1}{\sqrt{2}} \left(\begin{array}{ccc} 0&0&0\\0&0&1\\ 0&0&0
\end{array}\right),\quad E_{12} = [E_1,E_2]= \frac{1}{2} \left(\begin{array}{ccc} 0&0&1\\0&0&0\\ 0&0&0
\end{array}\right).
\ee
Correspondingly, the negative roots are
\be
\Theta_1=\frac{1}{\sqrt{2}} \left(\begin{array}{ccc} 0&0&0\\1&0&0\\ 0&0&0
\end{array}\right),\quad \Theta_2= \frac{1}{\sqrt{2}} \left(\begin{array}{ccc} 0&0&0\\0&0&0\\ 0&1&0
\end{array}\right),\quad \Theta_{12} = [\Theta_2,\Theta_1]= \frac{1}{2} \left(\begin{array}{ccc} 0&0&0\\0&0&0\\ 1&0&0
\end{array}\right).
\ee

Conversely, we denote the generators in the Chevalley basis by lower-case letters. The Cartan elements are
\be
\eta_1=\left(\begin{array}{ccc} 1&0&0\\0&-1&0\\ 0&0&0
\end{array}\right)\,,\qquad \eta_2=\left(\begin{array}{ccc} 0&0&0\\0&1&0\\ 0&0&-1
\end{array}\right)\,,
\ee
while we denote simple and highest roots as
\be
\eps_1=\left(\begin{array}{ccc} 0&1&0\\0&0&0\\ 0&0&0
\end{array}\right),\qquad \eps_2=\left(\begin{array}{ccc} 0&0&0\\0&0&1\\ 0&0&0
\end{array}\right),\qquad \eps_{12} = [\eps_1,\eps_2]=\left(\begin{array}{ccc} 0&0&1\\0&0&0\\ 0&0&0
\end{array}\right),
\ee
and the negative roots correspondingly as 
\be
\th_1=\left(\begin{array}{ccc} 0&0&0\\1&0&0\\ 0&0&0
\end{array}\right),\qquad \th_2=\left(\begin{array}{ccc} 0&0&0\\0&0&0\\ 0&1&0
\end{array}\right),\qquad \th_{12} = [\th_2,\th_1]=\left(\begin{array}{ccc} 0&0&0\\0&0&0\\ 1&0&0
\end{array}\right).
\ee


\section{BPS equations with defects}
\label{ap:defects}

For convenience, here we spell out in detail the notation concerning the defect-BPS equations used in the main text, based on \cite{Beasley:2008dc}. The setting we are interested in is a 7-brane stack hosting an 8d SYM, which is coupled to defects localised at the intersection with another 7-brane stack. Take $S$ and $S'$ to be these two 4-cycles intersecting in a complex curve $\Sigma \equiv S \cap S'$, which we take to be irreducible and smooth for simplicity. If we denote the two gauge groups as $G_S, G_{S'}$, the matter content of the theory can then be decomposed as
\begin{align}
\mr{ad}(G_\Sigma) = \mr{ad} (G_S) \oplus \mr{ad} (G_{S'}) \oplus \left( \bigoplus_j U_j \otimes U_j' \right), 
\end{align}
where the last part corresponds to additional matter localised on $\Sigma$ transforming in bifundamental representations $U,U'$ of the two gauge groups and $G_\Sigma$ denotes the enhanced-symmetry group found along this locus. In particular the defect theory contains a pair of complex scalars $(\sigma,\sigma^c)$ transforming as
\begin{subequations}
	\label{sections}
	\begin{align}
	\sigma &\in \Gamma \left( K_\Sigma^{1/2} \otimes \mc{U} \otimes \mc{U}' \right) \\
	\sigma^c &\in \Gamma \left( K_\Sigma^{1/2} \otimes \mc{U}^* \otimes (\mc{U}')^* \right),
	\end{align}
\end{subequations}
where we denoted by $\mc{U},\mc{U}'$ the vector bundles associated to $U, U'$, which are determined by restricting the principal bundles on the 7-brane stacks to $\Sigma$.

In the following we will denote by $\langle \cdot, \cdot\rangle_{\mc{U}}$  the natural product between $\mc{U}$ and its dual bundle $\mc{U}^*$ and accordingly for $\mc{U}'$. This product induces a map to the Lie-algebra $\mg{g}_S$ of $G_S$. If we denote the action of the generators of $\mg{g}_S$ in $\mc{U}$ by $T$, it is given by
\begin{align}
T: \mc{U}^* \times \mc{U} &\longrightarrow \mg{g}_S \label{generator}\\
(u, v) &\mapsto \langle T \cdot, \cdot \rangle_{\mc{U}} \nonumber.
\end{align}

Note, moreover that the bundles $\mc{U}, \mc{U'}$ and $K_\Sigma^{1/2}$ are all hermitian and therefore equipped with a metric
\begin{subequations}
	\label{metric}
	\begin{align}
	H&: \mc{U} \longrightarrow \bar{\mc{U}}^*\,, \\
	H'&: \mc{U}'\longrightarrow \bar{\mc{U}}'^*\,, \\
	h_\Sigma^{-1/2}&: K_\Sigma^{1/2} \longrightarrow \bar{K}_\Sigma^{-1/2}.
	\end{align}
\end{subequations}

With these maps at hand, we may now construct the product and moment map introduced in \ref{susy7d}. Recall that they are maps 
\begin{align}
\llangle \cdot , \cdot \rrangle_{\mg{g}_S}&: &\left( K_\Sigma^{1/2} \otimes \mc{U}^* \otimes (\mc{U}')^* \right) &&\times &&\left( K_\Sigma^{1/2} \otimes \mc{U} \otimes \mc{U}' \right) &&\longrightarrow &&K_\Sigma \otimes \mg{g}_S \label{holmom} \,, \\
\mu&: &\left( \bar{K}_\Sigma^{1/2} \otimes \bar{\mc{U}} \otimes \bar{\mc{U}'} \right) &&\times &&\left( K_\Sigma^{1/2} \otimes \mc{U} \otimes \mc{U}' \right) &&\longrightarrow &&\mg{g}_S \label{Dmom}\,,,
\end{align}
the first of which can now be composed out of the natural product of $\mc{U}'$ and \ref{generator} as
\begin{align}
\llangle \cdot , \cdot \rrangle_{\mg{g}_S} &= \langle T \cdot, \cdot \rangle_{\mc{U}} \otimes \langle \cdot, \cdot \rangle_{\mc{U}'}\,,
\end{align}
while the second also involves the hermitian bundle metrics $H,H', h_{\Sigma}^{-1/2}$ as 
\begin{align}
\mu:&=  \langle h_\Sigma^{-1/2} \cdot, \cdot \rangle_{K_\Sigma^{1/2}} \langle TH \cdot, \cdot \rangle_{\mc{U}} \langle H' \cdot, \cdot \rangle_{\mc{U}'}\,.
\end{align}
Locally, we may therefore write
\begin{align}
	\llangle\sigma^c, \sigma\rrangle_{\mg{g}_S} &= \sigma^{c\,A}_j (T^I)^j{}_{i} \sigma^{i}_A \, \mg{t}_I \,,\\
	\mu &= h_\Sigma^{-1/2} \left[(H^{\prime\, -1})^{\bar{A}B} \bar{\sigma}^{\bar{k}}_{\bar{A}} H_{\bar{k}j} (T^I)^j{}_{i} \sigma^i_B - H'_{B\bar{A}} \sigma^{c \, B}_i (H^{-1})^{j\bar{k}} (T^I)_{j}{}^i  \bar{\sigma}^{c \, \bar{A}}_{\bar{k}} \right] \mg{t}_I\,,
\end{align}
where we denoted $\mg{t}_I$ the generators of $\mg{g}_S = \mr{Lie}(G_S)$, $T^I$ the same generators in the representation $U$ with indices $i,j$, and we have used that $-(T^{I})^T$ are the generators in the conjugate representation $U'$. $A,B$, instead, are indices of $U'$, which are always contracted in such a way that we get singlets under $G_{S'}$. Note, that the above explicit expressions hold globally on $\Sigma$ in the case that both $\mc{U}$ and $\mc{U}'$ are split bundles.


\section{4d reduction and massive modes}
\label{ap:4d}

Section \ref{s:4d} discusses the four-dimensional picture related to 7-branes with defects, but focusing on the most relevant subcase and omitting many technical details. The purpose of this appendix is to give a more detailed description of this 4d picture. We have organised this appendix in the same way as section \ref{s:4d} to make the comparison as simple as possible. Before we begin with the physical analysis, let us discuss the different form-eigenbases of the Laplacian we will need for the computation as well as some mathematical conventions. Throughout the appendix we take the convention of summing over repeated indices, except for the dummy-index $\bullet$. Moreover, in some equations we separate zero-modes indicated with the index $i_0$ from non-zero modes indexed by $i$.

\subsection*{The spectrum of bulk KK modes}
Let us quickly review the notation we use with respect to Hodge star, scalar product and adjoint operators. We denote by $*$ the map
\begin{align}
*: \Omega^{(p,q)} &\longrightarrow \Omega^{(2-q,2-p)},
\end{align}
which induces a scalar product
\begin{align}
\left< \alpha, \beta \right> &\equiv \frac{1}{V_S} \int_S \alpha \w * \bar{\beta}\,,
\end{align}
and it is with respect to this scalar product that we define the adjoint differential operators
\begin{align}
\langle \pb \alpha, \beta \rangle &= \langle \alpha , \pb^\dagger \beta \rangle \\
\Rightarrow \pb^\dagger &= - * \p *.
\end{align}

Recall, that our T-brane example of the main text is given in $\mg{su}(2)$, such that the forms that appear are valued in three different bundles, corresponding to the three generators of $\mg{su}(2)$. We therefore denote by $\psi_3^I$ and $\psi_\pm^I$ these three $(0,1)$-form eigenbases of the Laplacian
\begin{align}
\Delta_{\pb_\bullet} \psi_\bullet^I \equiv -(l_\bullet^I)^2 \psi_\bullet^I \,,\label{1forms}
\end{align}
and accordingly the $(2,0)$-form bases as $\chi_3^A$ and $\chi_\pm^A$
\begin{align}
\Delta_{\pb_\bullet} \chi_\bullet^A = -(k_\bullet^A)^2 \chi_\bullet^A, \label{2forms}
\end{align}
where there is no summation over the repeating indices. Moreover, we take both bases to be orthornormal. That is 
\begin{align}
\delta^{AB} &= \frac{1}{V_S} \int_S \chi_\bullet^A \w \bar{\chi}_\bullet^B\, , \\
\delta^{IJ} &= \frac{1}{V_S} \int_S \psi_\bullet^I \w * \bar{\psi}_\bullet^J\, .
\end{align}
Recall the gauge covariant derivative and its Laplacian
\begin{align}
\pb_A^\dagger &= - * \p_A *\, , \\
\Delta_{\pb_A} &= \pb_A \pb_A^\dagger + \pb_A^\dagger \pb_A\, ,
\end{align}
and let us define its action on the one-form bases as
\begin{align}
\pb \psi_3^I &\equiv i \mu_{3\, A}^I \bar{\chi}_3^A\, , \label{deract1}\\
\left( \pb_A \psi^I \right)_\pm &\equiv i \mu_{\pm\, A}^I \bar{\chi}_\mp^A\, , \label{deract2}
\end{align}
Note, that this equation gives us a relation between the eigenvalues of the Laplacian for the two bases. Namely, by acting with the Laplacian on both sides of the equation we get
\begin{align}
\Rightarrow \pb \psi_3^I &\equiv i \frac{(k_3^A)^2}{(l_3^I)^2} \mu_{3\, A}^I \bar{\chi}_3^A \, ,
\label{lapl1}
\end{align}
such that for a given pair {$(I,A)$ eqs.\eqref{deract1} and \eqref{lapl1} may only be satisfied if either $\mu_{3\, A}^I = 0$ or $(k_3^A)^2 = (l_3^I)^2$.

As we will see below, the superpotential couples one- and two-forms in the Yukawa-couplings and therefore we will need to give relations between the $(0,1)$-form and the $(2,0)$-form basis, such that we define the set of constants $\Lambda$ as
\begin{align}
\Lambda^{IJA}_{3}&=\int_S \psi_+^I\wedge\psi_-^J\wedge\chi_3^A\, , \\
\Lambda^{IJA}_{+}&=\int_S \psi_3^I\wedge\psi_+^J\wedge\chi_3^A\, , \\
\Lambda^{IJA}_{-}&=\int_S \psi_3^I\wedge\psi_-^J\wedge\chi_3^A\, .
\end{align}

Finally, to integrate the 6d superpotential we need to introduce the following set of couplings:
\be
\Gamma^{I \kappa_1 \kappa_2}_{+} = \int_S \de_\Sigma\wedge (\sigma_1^c)^{\kappa_1} (\sigma_2)^{\kappa_2} \wedge \psi^I_+\, .
\ee

\subsection*{F-terms without defects}
We will start by computing the four dimension superpotential from 8d SYM and then in a second step compute the additional contributions induced by defects. Recall that
\begin{align}
W_S &= \int_S \mr{Tr}\,  \Phi \w \mb{F} \, . \label{8dsuperpot}
\end{align}
As in section \ref{s:4d}, we will work in the case of an $SU(2)$ split-bundle and are now interested to study infinitesimal fluctuations around the background $\langle \Phi\rangle = 0$ and $\mb{A} = A \, T_3$, such that $\mb{F} \w J = 0$. We denote the fluctuations by
\begin{align}
	\delta A^{(0,1)} &\equiv \left(\begin{matrix}
		a_3 & a_+ \\ a_- & - a_3
	\end{matrix}\right), &
	\delta \Phi &\equiv \left(\begin{matrix}
		v & m \\ p & -v
	\end{matrix}\right).
\end{align}
Let us now pass to four dimensions by expanding the modes in suitable basis. To this end, recall that the relevant fields transform as
\begin{align}
a_3 &\in \Omega^{0,1}(S,{\mc{O}}), \hspace{1cm} a_\pm \in \Omega^{0,1}(S,\mc{L}^{\pm 2}), \\
v &\in \Omega^{2,0}(S,\mc{O}), \hspace{1cm} m \in \Omega^{2,0}(S,\mc{L}^2), \hspace{1cm} p \in \Omega^{2,0}(S,\mc{L}^{-2}).
\end{align}
Each of these six spaces needs to be expanded in its own basis, defined in \eqref{1forms} and \eqref{2forms}:
\begin{align}
v &\equiv v_A \chi_3^A, & m &\equiv m_A \chi_+^A, & p &\equiv p_A \chi_-^A, & a_\bullet \equiv a_{\bullet I } \psi_\bullet^I.
\end{align}
Plugging this into \eqref{8dsuperpot} gives
\begin{align}
W_{\rm 4d} & = i \mu_{3\, A}^I v_A a_{3\, I} - 2i v_A a_{+\, I} a_{-\, J} \Lambda_{3}^{IJA} \label{4dsuperpot} \\
&+ i\mu_{-\, A}^I m_A a_{-\, I} + 2 im_A a_{3\, I} a_{-\, J} \Lambda_{-}^{IJA} + i\mu_{+\, A}^I p_A a_{+\, I} - 2 ip_A a_{3\, I} a_{+\, J} \Lambda_{+}^{IJA}\, . \nonumber
\end{align}
We may easily read off the F-terms for $v,m$ and $p$ from each line, while those for $a_\bullet$ are 
\begin{align}
0 &= v_A \mu_{3\, A}^I + m_A a_{-J} \Lambda_{-}^{IJA} - p_A a_{+\, J} \Lambda_{+}^{IJA}\, , \label{vA}\\
0 &= m_A \mu_{-\, A}^I + 4 v_A a_{+\, J} \Lambda_{3}^{IJA} - 2 m_A a_{3\, j} \Lambda_{-}^{IJA}\, , \label{m} \\
0 &= p_A \mu_{+\, A}^I - 4 v_A a_{-\, J} \Lambda_{3}^{IJA} + 2 p_A a_{3\, j} \Lambda_{+}^{IJA}\, , \label{p}
\end{align}
where we have used that $\Lambda^{JIA} = - \Lambda^{IJA}$. Note that $\mu_{\bullet\,  A}^I$ vanishes if either $A$ or $I$ correspond to harmonic forms. Restricting to the remaining forms, $\mu$ is in fact invertible and we may express the F-terms for $v,m$ and $p$} as
\begin{align}
\mu_{3\, I}^A a_{3\, I} &= 2 a_{+\, K} a_{-\, J} \Lambda_{3}^{KJA} \, ,\label{a3}\\
0 &= \left( \mu_{-A}^I - 2 a_{3\, J} \Lambda_{-}^{IJA} \right) a_{-\, I}\, , \label{a-A}\\
0 &= \left( \mu_{+\, A}^I + 2 a_{3\, J} \Lambda_{+}^{IJA} \right) a_{+\, I}\, . \label{a+A}
\end{align}

\paragraph{Massive modes}
We can see from eq. \eqref{a3} - \eqref{a+A} that one solution is given by $a_3 = a_\pm = 0$ and from \eqref{vA} - \eqref{p} that this allows for $v = p = m = 0$. This is the solution we are after. In general there are no other solutions.\footnote{To be more precise: More solutions exist, if either we have $\mu_{+,A}^I = A \, \mu_{-,A}^I$ and $\Lambda_{+}^{IJA} = -A \, \Lambda_{-}^{IJA}$ for some $A \in \mb{C}$ or that $\mr{supp}(\Lambda_+) \subseteq \ker(\Lambda_-)$ and vice-versa.}

\paragraph{Zero modes}
Let us focus for a moment on zero-modes, for which all the $\mu$'s vanish. Let us moreover assume that $a_3$ contains no zero-modes because $S$ is simply-connected so in particular the e.o.m with respect to it does not exist for zero-modes. Then the zero-modes need to satisfy
\begin{align}
0 &= a_{+\, I} a_{-\, J} \Lambda_{3}^{IJA}\, ,  \\
0 &= v_A a_{-\, J} \Lambda_{3}^{IJA}\, , \\
0 &= v_A a_{+\, J} \Lambda_{3}^{IJA}\, .
\end{align}
Apart from pathological cases, this implies that for each set of indices two of the three fields must vanish.\footnote{If for every $A$, the matrices $\Lambda_{3}^{IJA}$ regarded as a map to $\mb{C}$ have the \underline{same} non-trivial kernel, then these conditions do not imply the vanishing of the individual fields in the above equations. That is, the double sum over $IJ$ might allow for cancellations.}

\subsection*{F-terms with defects}
Let us now compute the additional defect contributions to the superpotential. The six-dimensional defect superpotential is given by
\be\label{defsuperpot}
	W_\Sigma = \int_\Sigma \llangle \sigma^c, \pb_{\mb{A}} \sigma \rrangle  = \int_\Sigma \llangle \sigma^c, \pb_{\langle\mb{A}\rangle} \sigma - i \delta \mb{A} (\sigma) \rrangle\, ,
\ee
where by $\delta \mb{A} (\sigma)$ we denoted the action of $\delta \mb{A}$ in the fundamental representation of $\mg{su}(2)$ on $\sigma$. Recall from \eqref{def1} and \eqref{def2} that the fields $\sigma^c, \sigma$ transform as sections of $K_\Sigma^{1/2}$, whereas the product $\sigma^c \sigma$ transforms as a section of $K_\Sigma$, such that it is much more convenient to expand the product of the two fields in our previous bases $\{ \psi_3, \psi_+, \psi_- \}$, than to expand the individual fields\footnote{At the cost of more notation, one might also expand the fields individually and then define a set of additional coefficients, that relate the basis of $0$-forms valued in $K_\Sigma^{1/2} \otimes \mc{L}^\pm$ to those of $(1,0)$-forms valued in $\mc{L}^{\pm }$ and $\mc{O}$.}
\begin{align}
\sigma_1^c \sigma_1 &= \left(\sigma_1^c \sigma_1\right)_{i_0} \bar{\psi}_3^{i_0}\, , \hspace{1cm} \sigma_2^c \sigma_2 = \left(\sigma_2^c \sigma_2\right)_{i_0} \bar{\psi}_3^{i_0}\, ,\\
\sigma_1^c \sigma_2 &= \left(\sigma_1^c \sigma_2\right)_{i_0} \bar{\psi}_+^{i_0}\, , \hspace{1cm} \sigma_2^c \sigma_1 = \left(\sigma_2^c \sigma_1\right)_{i_0} \bar{\psi}_-^{i_0}\, . \nonumber
\end{align}
Plugging all of this into \eqref{defsuperpot}, gives
\begin{align}
W_\Sigma &=  i \sigma_{2 \kappa_1}^c \sigma_{2 \kappa_2} \Gamma^{J \kappa_1 \kappa_2} a_{3\, J} - i \sigma_{1 \kappa_1}^c \sigma_{1 \kappa_2} \Gamma^{J \kappa_1 \kappa_2} a_{3\, J}  \label{4ddefsuperpot} \\
&- i \sigma_{1 \kappa_1}^c \sigma_{2 \kappa_2} \Gamma^{J \kappa_1 \kappa_2} a_{+\, J} - i \sigma_{2 \kappa_1}^c \sigma_{1 \kappa_2} \Gamma^{J \kappa_1 \kappa_2} a_{-\, J}. \nonumber
\end{align}

With both superpotential contributions eqs.\eqref{4dsuperpot} and \eqref{4ddefsuperpot} at hand, we may now compute the equations of motion. Those for the component fields of $\delta \Phi$ can be easily read off from \eqref{4dsuperpot} and do not depend on the defect fields. On the other hand those for the components of $\delta \mb{A}$ are given by 
\begin{subequations}
	\begin{align}
		\Gamma^{I \kappa_1 \kappa_2} \left(\sigma_{1 \kappa_1}^c \sigma_{1 \kappa_2} - \sigma_{2 \kappa_1}^c \sigma_{2 \kappa_2} \right) &= v_A \mu_{3\, A}^I + m_A a_{-\, J} \Lambda_{-}^{IJA} - p_A a_{+\, J} \Lambda_{+}^{IJA}\, , \\
		\Gamma^{I \kappa_1 \kappa_2} \sigma_{2 \kappa_1}^c \sigma_{1 \kappa_2} &= m_A \mu_{-\, A}^I + 2 v_A a_{+\, J} \Lambda_{3}^{IJA} - m_A a_{3\, j} \Lambda_{-}^{IJA}\, , \\
		\Gamma^{I \kappa_1 \kappa_2} \sigma_{1 \kappa_1}^c \sigma_{2 \kappa_2} &= p_A \mu_{+\, A}^I -2 v_A a_{-\, J} \Lambda_{3}^{IJA} + p_A a_{3\, j} \Lambda_{+}^{IJA}\, ,
	\end{align}
\end{subequations}
while those for the defect fields themselves are given as
\begin{subequations}
	\begin{align}
		0 &=  \Gamma^{J \kappa_1 \kappa_2} \left( \sigma_{1 \kappa_1}^c a_{3\, J} + \sigma_{2 \kappa_1}^c a_{-\, J} \right) \\
		0 &=  \Gamma^{J \kappa_1 \kappa_2} \left( \sigma_{2 \kappa_1}^c a_{3\, J} - \sigma_{1 \kappa_1}^c a_{+\, J}  \right) \\
		0 &=  \Gamma^{J \kappa_1 \kappa_2} \left( \sigma_{1 \kappa_2} a_{3\, J} + \sigma_{2 \kappa_2} a_{+\, J} \right) \\
		0 &=  \Gamma^{J \kappa_1 \kappa_2} \left( \sigma_{2 \kappa_2} a_{3\, J} - \sigma_{1 \kappa_2} a_{-\, J} \right).
	\end{align}
\end{subequations}

Let us now try to solve them, at least partially. We proceed separately for zero modes and massive modes.

\paragraph{Massive modes}
For the massive modes the story is more interesting. First, under the same caveats as for the non-defect case, we have $a_3 = a_\pm = 0$ --- so we may have only non-trivial vevs in the zero-mode part of the $a_\pm$ modes. In this notation the F-terms for the massive modes in $v,m,p$ read
\begin{subequations}
	\begin{align}
		v_A &= \left(\mu_{3\, A}^I\right)^{-1} \Gamma^{I \kappa_1 \kappa_2} \left( \sigma_{1 \kappa_1}^c \sigma_{1 \kappa_2} - \sigma_{2 \kappa_1}^c \sigma_{2 \kappa_2} \right) \\
		&- \left(\mu_{3\, A}^I\right)^{-1} \left( m_B a_{-\, j_0} \Lambda_{-}^{I j_0B} - p_B a_{+\, j_0} \Lambda_{+}^{I j_0B} \right)\, , \nonumber\\
		m_A &= \left(\mu_{+\, A}^I\right)^{-1} \Gamma^{I \kappa_1 \kappa_2} \sigma_{2 \kappa_1}^c \sigma_{1 \kappa_2} - 2 \left(\mu_{-\, A}^I\right)^{-1} v_B a_{+\, j_0} \Lambda_{3}^{I j_0B}\, , \label{4dmeq} \\
		p_A &= \left(\mu_{-, A}^I \right)^{-1} \Gamma^{I \kappa_1 \kappa_2} \sigma_{1 \kappa_1}^c \sigma_{2 \kappa_2} + 2 \left(\mu_{+\, A}^I\right)^{-1} v_B a_{-\, j_0} \Lambda_{3}^{I j_0B}\, .
	\end{align}
\end{subequations}
These are the generalised version of the relations found in eqs.\eqref{finalEq}.

\paragraph{Zero-modes}
Let us focus for a moment on zero-modes, that is all the $\mu$'s vanish. Let us moreover assume that $a_3$ contains no zero-modes because $S$ is simply-connected so in particular the e.o.m with respect to it does not exist for zero-modes. Then the zero-modes need to satisfy
\begin{subequations}
	\begin{align}
		0 &= a_{+k_0} a_{-j_0} \Lambda_{3}^{k_0 j_0 a_0} \\
		\Gamma^{i_0 \kappa_1 \kappa_2} \sigma_{2 \kappa_1}^c \sigma_{1 \kappa_2} &= 2 v_{a_0} a_{+j_0} \Lambda_{3}^{i_0 j_0 a_0} \\
		\Gamma^{i_0 \kappa_1 \kappa_2} \sigma_{1 \kappa_1}^c \sigma_{2 \kappa_2} &= -2 v_{a_0} a_{-j_0} \Lambda_{3}^{i_0 j_0 a_0} \label{4dzm}.
	\end{align}
\end{subequations}
Where we have used that the massive modes for $a_\pm$ have been dynamically set to zero, and that the Yukawa couplings of the form $\Lam^{i_0j_0 A}$ with two massless modes of $\mathbb{A}$ and one massive mode of $\Phi$ vanish identically. The same comments regarding a non-trivial kernel of the $\Lambda$'s as in the case without defects also apply here. The above equations generalise the F-term constraints found in eqs \eqref{FernandoConstraint}.

\end{document}